\pdfoutput=1  

\documentclass[preprint, nopreprintline, times, 10pt]{elsarticle}

\usepackage{amsmath}  
\usepackage[bookmarks,colorlinks]{hyperref}  
\usepackage{geometry}
\usepackage{txfonts}
\usepackage[caption=false,font=footnotesize,labelfont=sf,textfont=sf]{subfig}
\usepackage{graphicx}
\usepackage{amssymb}
\usepackage{amsfonts}  
\usepackage{booktabs}  
\usepackage[linesnumbered,ruled,noend]{algorithm2e}  
\usepackage{soul}  
\usepackage{color, xcolor}  
\usepackage{xspace}  
\usepackage{url}  
\usepackage{xparse}  
\usepackage[skins]{tcolorbox}

\graphicspath{ {assets/figure/}, {assets/image/} }
\DeclareGraphicsExtensions{.png, .pdf, .jpg}
\newtheorem{theorem}{\bf Theorem}
\newtheorem{corollary}{\bf Corollary}
\newtheorem{lemma}{\bf Lemma}
\newdefinition{definition}{\bf Definition}
\newproof{proof}{\bf Proof}
\definecolor{blk}{HTML}{000000} 
\definecolor{b1}{HTML}{007FFF} 
\definecolor{g1}{HTML}{42CB01} 
\definecolor{p1}{HTML}{6666cc} 
\definecolor{y1}{HTML}{F1BE12} 
\tcbuselibrary{breakable, listings}
\newtcolorbox[
    auto counter, 
    number within=section, 
    list type=section, 
    list inside=toc]{questionbox}[1]{
        colback=white, 
        colframe=blk, 
        colbacktitle=blk, 
        coltitle=white, 
        fonttitle=\bfseries, 
        title={Comment \thetcbcounter}, 
        list entry={Comment \thetcbcounter\quad}, 
        breakable, 
        before upper={\parindent10pt\noindent}, 
        left = 1 mm, 
        right = 1 mm, 
        top = 1 mm, 
        bottom = 1 mm,
        right skip = 0\textwidth,
        enhanced,drop fuzzy shadow,  
}

\newif\ifaddfigure
\addfiguretrue

\NewDocumentCommand \sigfd     {}{\varepsilon} 
\NewDocumentCommand \dif       {}{\mathop{}\!\mathrm{d}} 
\NewDocumentCommand \numedgetype{}{L}
\NewDocumentCommand \edgeij    {}{\edge(ij)}
\NewDocumentCommand \randsubset{}{\mathcal{C}}
\NewDocumentCommand \candset   {}{C}
\NewDocumentCommand \visitedset{}{Z}
\NewDocumentCommand \nodei     {}{\node(i)}
\NewDocumentCommand \nodej     {}{\node(j)}
\NewDocumentCommand \source    {}{\node(s)}
\NewDocumentCommand \pathsij   {}{\paths(ij)}
\NewDocumentCommand \layers    {}{l} 
\NewDocumentCommand \recurrence{}{\mathcal{R}}
\NewDocumentCommand \modelparam{}{\Theta} 
\NewDocumentCommand \prob      {}{P}
\NewDocumentCommand \uniprob   {}{p} 
\NewDocumentCommand \singlepath{}{m}
\NewDocumentCommand \Q         {}{q} 
\NewDocumentCommand \Qt        {}{\mathbb{Q}} 
\NewDocumentCommand \ie        {}{\text{i}.\text{e}.\xspace}
\NewDocumentCommand \eg        {}{\text{e}.\text{g}.\xspace}

\NewDocumentCommand \sdtd        {}{\text{s.t.}\,}

\NewDocumentCommand \period    {}{{.}\xspace}
\NewDocumentCommand \framework {}{IGNP\xspace}
\NewDocumentCommand \frameworkname{}{influential graph neural predictor\xspace}  

\NewDocumentCommand \wiki      {}{\textit{WikiData}\xspace}
\NewDocumentCommand \ppi       {}{\textit{PPI}\xspace}
\NewDocumentCommand \twitter   {}{\textit{Twitter}\xspace}
\NewDocumentCommand \dblp      {}{\textit{DBLP}\xspace}
\NewDocumentCommand \blogcatalog{}{\textit{BlogCatalog}\xspace}
\NewDocumentCommand \bigO      {}{O}
\NewDocumentCommand \Lone      {}{L^1} 
\NewDocumentCommand \Ltwo      {}{L^2} 
\NewDocumentCommand \mf        {}{f} 
\NewDocumentCommand \argmin    {}{\mathop{\arg\min}}
\NewDocumentCommand \lastepoch {}{\alpha} 
\NewDocumentCommand \evalthsh  {}{\xi} 
\NewDocumentCommand \ngbat     {}{b} 
\NewDocumentCommand \tsp       {}{\ts[\prime]} 
\NewDocumentCommand \E         {}{E} 
\NewDocumentCommand \normal    {}{\textit{susceptible}\xspace}
\NewDocumentCommand \influential{}{\textit{infectious}\xspace}
\NewDocumentCommand \influenced{}{\textit{recovered}\xspace}
\NewDocumentCommand \noticed   {}{\textit{noticed}\xspace}
\NewDocumentCommand \graph     { D[]{} D(){} }{\mathcal{G}^{#1}_{#2}}
\NewDocumentCommand \hall      { D[]{} D(){} }{\mathcal{H}_{#2}}
\NewDocumentCommand \hin       { D[]{} D(){} }{\mathcal{H}^{\text{in}}_{#2}}
\NewDocumentCommand \hout      { D[]{} D(){} }{\mathcal{H}^{\text{out}}_{#2}}
\NewDocumentCommand \propEvt   { D[]{} D(){} }{\mathbb{X}^{#1}_{#2}} 
\NewDocumentCommand \loss      { D[]{} D(){} }{\mathcal{L}^{#1}_{#2}}
\NewDocumentCommand \border    { D[]{} D(){} }{\mathcal{B}^{#1}_{#2}}
\NewDocumentCommand \adj       { D[]{} D(){} }{\mathbf{A}^{#1}_{#2}}
\NewDocumentCommand \paths     { D[]{} D(){} }{\mathcal{P}^{#1}_{#2}}
\NewDocumentCommand \neighbor  { D[]{} D(){} }{\Gamma^{#1}_{#2}}
\NewDocumentCommand \recovery  { D[]{} D(){} }{\tau^{#1}_{#2}}
\NewDocumentCommand \ts        { D[]{} D(){} }{{t}^{#1}_{#2}} 
\NewDocumentCommand \mInf      { D[]{} D(){} }{\upsilon^{#1}_{#2}}
\NewDocumentCommand \mS        { D[]{} D(){} }{S^{#1}_{#2}} 
\NewDocumentCommand \ms        { D[]{} D(){} }{s^{#1}_{#2}}
\NewDocumentCommand \mI        { D[]{} D(){} }{I^{#1}_{#2}}
\NewDocumentCommand \mi        { D[]{} D(){} }{i^{#1}_{#2}}
\NewDocumentCommand \mR        { D[]{} D(){} }{R^{#1}_{#2}}
\NewDocumentCommand \mr        { D[]{} D(){} }{r^{#1}_{#2}}
\NewDocumentCommand \mU        { D[]{} D(){} }{U^{#1}_{#2}}
\NewDocumentCommand \repknl    { D[]{} D(){} }{{M}^{#1}_{#2}} 
\NewDocumentCommand \nborder   { D[]{} D(){} }{{n}^{#1}_{#2}} 
\NewDocumentCommand \evals     { D[]{} D(){} }{{\eta}^{#1}_{#2}}
\NewDocumentCommand \edge      { D[]{} D(){} }{e^{#1}_{#2}}
\NewDocumentCommand \node      { D[]{} D(){} }{\mathbf{x}^{#1}_{#2}}
\NewDocumentCommand \nodeset   { D[]{} D(){} }{\mathcal{X}^{#1}_{#2}}
\NewDocumentCommand \edgefeat  { D[]{} D(){} }{\text{Dist}^{#1}_{#2}}
\NewDocumentCommand \gsparsity { D[]{} D(){} }{\rho^{#1}_{#2}}
\NewDocumentCommand \vsgraph   { D[]{} D(){} }{\mathcal{S}^{#1}_{#2}} 
\NewDocumentCommand \noticedset{ D(){s} }{{\mathcal{X}^{\text{Noticed}}_{#1}}}
\NewDocumentCommand \fsgraph   { D(){} }{\mathcal{S}^{F}_{#1}} 
\NewDocumentCommand \lsgraph   { D(){} }{\mathcal{S}^{L}_{#1}} 
\NewDocumentCommand \cbkt      { m D(){} }{{\{ {#1} \}}_{#2}} 
\NewDocumentCommand \real      { D[]{} }{\mathbb{R}^{#1}} 
\NewDocumentCommand \netweight { D[]{} }{\mathbf{W}^{#1}} 
\NewDocumentCommand \nodefeat  { D[]{} }{\mathcal{X}^{#1}} 
\NewDocumentCommand \diagmat   { D[]{} }{\mathbf{D}^{#1}}
\NewDocumentCommand \edgeset   { D[]{} D(){} }{\mathcal{E}^{#1}_{#2}}
\NewDocumentCommand \realdim   { D[]{} }{{d}^{#1}} 
\NewDocumentCommand \rbkt      { m D[]{} }{{( {#1} )}^{#2}} 
\NewDocumentCommand \bbkt      { m D[]{} }{{[ {#1} ]}^{#2}} 
\NewDocumentCommand \abkt      { m }{\langle {#1} \rangle} 
\NewDocumentCommand \dabs      { m D[]{} D(){} }{|| {#1} ||^{#2}_{#3}} 
\NewDocumentCommand \eqdabs    { m D[]{} D(){} }{\left\lVert {#1} \right\rVert^{#2}_{#3}} 
\NewDocumentCommand \eqcbkt    { m }{\left\{ {#1} \right\}} 
\NewDocumentCommand \eqlcbkt   { m }{\left\{ {#1} \right.} 
\NewDocumentCommand \eqrbkt    { m D[]{} }{{\left( {#1} \right)}^{#2}} 
\NewDocumentCommand \eqbbkt    { m }{\left[ {#1} \right]} 
\NewDocumentCommand \eqabs     { m D[]{} }{{\left\vert {#1} \right\vert}^{#2}}
\NewDocumentCommand \abs       { m D[]{} }{{\vert {#1} \vert}^{#2}}
\NewDocumentCommand \infprob   { m }{P_{e_{ {#1} }}} 
\NewDocumentCommand \lpfunc    { m }{F_{lp}( {#1} )} 
\NewDocumentCommand \featfunc  { m }{F_{feat}( {#1} )} 
\NewDocumentCommand \figref    { m }{Fig.~\ref{#1}}
\NewDocumentCommand \tabref    { m }{Table~\ref{#1}}
\NewDocumentCommand \eqaref    { m }{Eq.~\eqref{#1}}
\NewDocumentCommand \defref    { m }{Definition~\ref{#1}}
\NewDocumentCommand \theref    { m }{Theorem~\ref{#1}}
\NewDocumentCommand \therefs   { m m }{Theorems~\ref{#1} and~\ref{#2}}
\NewDocumentCommand \cororef   { m }{Corollary~\ref{#1}}
\NewDocumentCommand \lemref    { m }{Lemma~\ref{#1}}
\NewDocumentCommand \lemrefs   { m m }{Lemmas~\ref{#1} and~\ref{#2}}
\NewDocumentCommand \secref    { m }{Section~\ref{#1}}
\NewDocumentCommand \appref    { m }{\ref{#1}}
\NewDocumentCommand \agoref    { m }{Algorithm~\ref{#1}}
\NewDocumentCommand \parahead  { m }{\textit{#1}}
\NewDocumentCommand \ct        { m }{\textbf{#1}}
\NewDocumentCommand \alg       { D[]{tb} m m }{  
    \begin{algorithm}[{#1}]
        \DontPrintSemicolon
        \SetKwInOut{Input}{\bfseries Input}
        \SetKwInOut{Output}{\bfseries Output}
        \caption{{#2}}
        {#3}
    \end{algorithm}
}
\NewDocumentCommand \normalfig { D[]{tb} m }{
    \ifaddfigure
    \begin{figure}[{#1}]
        \centering
        {#2}
    \end{figure}
    \fi
}
\NewDocumentCommand \widefig   { D[]{t} m }{
    \ifaddfigure
        \begin{figure*}[{#1}]
            \centering
            {#2}
        \end{figure*}
    \fi
}
\NewDocumentCommand \normaltab { D[]{tb} m }{
    \begin{table}[{#1}]
        \centering
        {#2}
    \end{table}
}
\NewDocumentCommand \widetab   { D[]{t} m }{
    \begin{table*}[{#1}]
        \centering
        {#2}
    \end{table*}
}
\NewDocumentCommand \defi      { D[]{obj} m }{
    \begin{definition}[{#1}]
        {#2}
    \end{definition}
}
\NewDocumentCommand \lemm      { D[]{obj} m }{
    \begin{lemma}[{#1}]
        \itshape
        {#2}
    \end{lemma}
}
\NewDocumentCommand \theo      { m }{
    \begin{theorem}
        \itshape
        {#1}
    \end{theorem}
}
\NewDocumentCommand \zcoro     { m }{
    \begin{corollary}
        \itshape
        {#1}
    \end{corollary}
}
\NewDocumentCommand \pf        { m }{
    \begin{proof}
        {#1}
        \hfill 
    \end{proof}
}

\newcommand{\nop}[1]{}

\NewDocumentCommand \hlt       { D[]{b1} m }{\textcolor{#1}{#2}}

\begin{document}
    \begin{frontmatter}

\title{Link prediction on multi-relational graphs from \\ an influence propagation perspective\tnoteref{t1}}

\tnotetext[t1]{Accepted manuscript. Published in \textit{Pattern Recognition}, article 114039, 2026. Formal publication: \url{https://doi.org/10.1016/j.patcog.2026.114039}\\
\copyright~2026. This manuscript version is made available under the CC-BY-NC-ND 4.0 license \url{https://creativecommons.org/licenses/by-nc-nd/4.0/}}

\author[1]{Zidu~Yin}
\ead{zidu.yin@ynnu.edu.cn}
\author[2]{Yuankai~Qi}
\author[3]{Dong~Gong}
\author[5]{Ehsan~Abbasnejad}
\author[4]{Kun~Yue\texorpdfstring{\corref{cor}}{*}}
\ead{kyue@ynu.edu.cn}
\author[5]{Javen~Qinfeng~Shi}

\cortext[cor]{Corresponding author}

\address[1]{
    School of Information Science and Technology, Yunnan Normal University, Kunming, China
}

\address[2]{
    School of Computing, Macquarie University, Sydney, Australia
}

\address[3]{
    School of Computer Science and Engineering, The University of New South Wales, Sydney, Australia
}

\address[4]{
    School of  Information Science and Engineering, Yunnan University, Kunming, China
}

\address[5]{
    School of Computer and Mathematical Sciences, Adelaide University, Adelaide, Australia
}



\begin{abstract}
Predicting the existence and type of links (edges) between nodes in a multi-relational graph is key for applications from social interaction prediction to knowledge relationship identification.
Enhancing local features with relevant global information is crucial for accurate link prediction, yet it remains challenging.
We address this by modeling the relationship between node pairs as node influence.
That is, whether the node influence can be propagated and what type of influence is propagated indicates where and what type the edge is, which will be the most relevant local and global information to predict the edges.
To this end, we extend the Susceptible-Infectious-Recovered (SIR) epidemic model to capture the influence propagation of nodes on a large scale through sub-graph structures. 
Subsequently, these sub-graphs are compressed using virtual edges, thereby substantially reducing the computation associated with utilizing the global graph structure.
Finally, we propose the \textbf{I}nfluential \textbf{G}raph \textbf{N}eural \textbf{P}redictor, referred to as \framework ,\footnote{The code is available at \url{https://github.com/coodest/gn_link_prediction}.} a link prediction framework guided by influence propagation.
Extensive experiments demonstrate the superiority of the proposed method, which outperforms strong baselines by a large margin on the widely used and real-world datasets.
\end{abstract}

    \end{frontmatter}

\section{Introduction}\label{sec:introduction}

Link prediction, \ie to predict the edge existence~\cite{DBLP:journals/csur/MartinezBT17} and types~\cite{DBLP:journals/tkdd/WangLH24}, on multi-relational graphs is foundational for a wide range of applications including personalized recommendation~\cite{DBLP:journals/pr/GongZZZMZDT24}, automatic completion of knowledge graphs~\cite{DBLP:journals/eswa/WangLLWLJPL24}, and relationship reconstruction in protein interaction networks~\cite{DBLP:journals/tcbb/Altuntas23}.
The edge to predict is closely related to its neighbors' edges, as evidenced by numerous studies~\cite{zhou2009predicting,DBLP:conf/kdd/RibeiroSF17}.
Contemporary studies demonstrate that the existence and types of edges are not solely determined by their proximate neighbors, since they may also be globally affected by long-range dependencies~\cite{DBLP:conf/iclr/LiaoBTGUZ18}.
The global information is not supported by many methods~\cite{DBLP:conf/nips/ZhangC18}, due to high computational costs.
In addition, existing methods may introduce irrelevant information~\cite{DBLP:conf/kdd/ChiangLSLBH19} and ignore critical features~\cite{DBLP:conf/nips/HamiltonYL17} while processing global information.
Therefore, the main challenge of multi-relational graph link prediction is how to enhance local features by precisely finding and efficiently utilizing the relevant global information.

\normalfig[t]{
\centerline{\includegraphics[width=\textwidth]{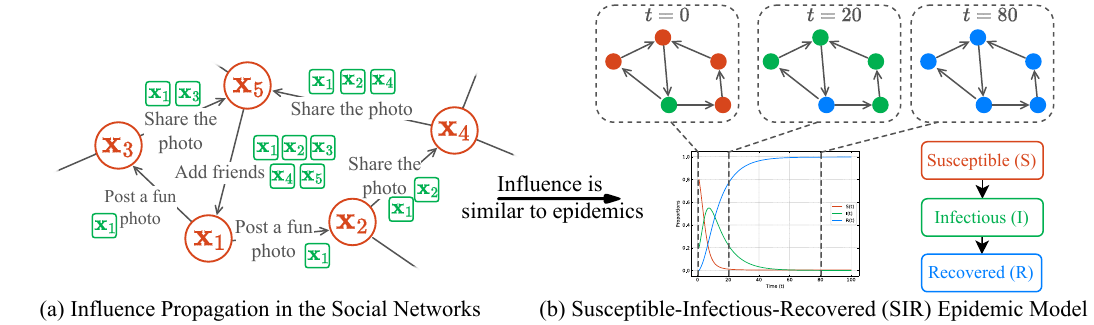}}
\caption{
An example of the influence propagation and epidemic propagation. (a) Influence propagation in the social networks. 
Red cycles, edge types, and green rectangles denote the user, the action, and the influence exerted by a specific user on the respective edge. (b) The SIR epidemic propagation model. The proportional changes are shown, and different colors indicate different types of nodes during the epidemic.
}
\label{fig:the-influence-propagation}
}

To consider as much relevant information as possible for link prediction, in this paper, we propose a novel propagation model to encapsulate the relevant local and global information into a sub-graph to train the GNN.

Through an influence propagation perspective, the influence, originating from the analysis of social networks, of the source node is the effect to make/alter edges of its neighbors with a specific type~\cite{DBLP:conf/wims/CercelT14}, which will propagate to different nodes within a range of the influence chain~\cite{DBLP:journals/jnca/PengZCYNJ18}.
An example of influence propagation on social networks is shown in \figref{fig:the-influence-propagation}(a).
The edges from $\node(1)$ lead to the edges of nearby nodes as chains of reaction, and some chains eventually go back to $\node(1)$ ($\node(1)$ - $\node(2)$ - $\node(4)$ - $\node(5)$ - $\node(1)$ and $\node(1)$ - $\node(3)$- $\node(5)$ - $\node(1)$ are two paths propagating the influence of $\node(1)$), \ie \textit{influence propagation}.
This means the source node $\node(1)$ externally impacts its surrounding nodes and internally assimilates responses from neighboring nodes, so the influence will subsequently be propagated within a range.
Additionally, whether the node influence can be propagated and what type of influence is propagated indicates the edge existence and type.
For instance, the edge between $\node(4)$ and $\node(5)$ is generated due to the overlapped influence of $\node(1)$, $\node(2)$ and $\node(4)$.
Therefore, edges can be predicted via node influence.

From the present example, the outward influence propagation demonstrates the analogous nature between the node influence and the epidemic outbreaks, since both of them change the surroundings.
The process of epidemic outbreaks can be precisely simulated and replicated, thereby enabling the tracking of node influence as well.
Specifically, we observe that in the susceptible-infectious-recovered (SIR) epidemic propagation model~\cite{DBLP:journals/amc/HarkoLM14}, the three types of nodes --- susceptible, infectious, and recovered --- are dynamically interchanged. This dynamic interchange helps formulate and track the outward spread of influence from the source until a stable state is reached, which marks the end of the outward influence propagation, as shown in \figref{fig:the-influence-propagation}(b).
Consequently, we use the SIR model to enclose the outward influence into the sub-graph.

Furthermore, it is imperative to consider the nodes' inward influence propagation that propagates from a broader range encompassing crucial global information.
Given this, we further expand the SIR model by introducing an inward propagation module to search the global responses back to their source.
Thus, mutual influence propagation paths in both directions should be established in the sub-graph, as they provide all the \textit{relevant information} for predicting the existence and types of edges.

More importantly, direct use of global influence within the sub-graph is nearly impossible due to the unaffordable computational cost, so we compress the sub-graph for a lower computational cost.
Inspired by the mean-field theory, which uses the averaged effect to represent the complex group effects~\cite{pastor2015epidemic}, we use virtual edges to virtually propagate the inward influences to the source within a sub-graph, thereby forming the \textit{virtual} sub-graph.
The virtual edge compresses global long-distance dependencies to minimize the sub-graph size, thereby reducing the computational costs of model training.
Experiments show that it greatly reduces the training cost without sacrificing much performance.

In summary, local node features are enhanced with relevant global information precisely identified by extending the SIR model, and we incorporate the enhanced features into the proposed virtual sub-graphs to efficiently train our \textbf{I}nfluential \textbf{G}raph \textbf{N}eural \textbf{P}redictor, namely \framework, for link prediction on multi-relational graphs.
Our contributions are:
\begin{enumerate}
    \item We extend the SIR propagation model to find more relevant local and global node information to enrich the data for better link prediction.
    \item We further compress inward influences with virtual edges to significantly reduce the training cost without significant performance degradation.
    \item We implement an effective link prediction framework \framework, which achieves the best accuracy compared with baselines from different categories on different datasets by processing both local and global node influences efficiently.
\end{enumerate}

The rest of the paper is organized as follows.
\secref{sec:related-work} briefly reviews related works.
\secref{sec:overview} provides an overview of the methodology and establishes the problem to be addressed.
\secref{sec:influence-propagation-theory} introduces the extended SIR propagation model to find the sub-graphs of influence propagation.
\secref{sec:the-framework} elucidates the \framework framework, which leverages the sub-graphs derived in \secref{sec:influence-propagation-theory}.
In \secref{sec:experiments} we present extensive evaluations of the proposed method.
The final section concludes this work.

\section{Related work}\label{sec:related-work}

\subsection{Link prediction}
Most existing link prediction methods are from similarity metrics, latent feature methods, and GNN-based methods~\cite{DBLP:journals/csur/MartinezBT17}, with comprehensive benchmarks further examining their evaluation protocols~\cite{NEURIPS2023_0be50b45}.

Early attempts use manually crafted similarity functions, such as Adamic Adar~\cite{DBLP:journals/socnet/AdamicA03}, Resource Allocation~\cite{zhou2009predicting}, and entropy-enhanced common neighbors~\cite{zhou2025common}, to leverage local topology for edge prediction, though they generalize poorly across diverse network structures.
Decomposed eigenvalues and eigenvectors~\cite{DBLP:journals/pr/Agibetov23} from the adjacency matrix~\cite{DBLP:conf/icdm/YanY23} have also been commonly employed for edge prediction.

Instead of directly using the topology data, many methods embed topological features as vectors into a latent space for broader applications.
LINE~\cite{DBLP:conf/www/TangQWZYM15} takes neighboring nodes within 2 hops as the context to acquire the node features. 
Node2vec~\cite{DBLP:conf/kdd/GroverL16} and Struc2vec~\cite{DBLP:conf/kdd/RibeiroSF17} produce a local context via the biased random walk to learn the node embeddings. 
HARP~\cite{DBLP:conf/aaai/ChenPHS18} uses an expandable context to enhance the embedding (e.g., Node2vec) iteratively with global information. 
Ripple2Vec~\cite{Luo2022ripple2vecNE} recursively aggregates neighbors via attention, letting influence ripple outward to encode multi-hop structure.
Generally, embeddings from latent feature methods store more useful features compared with the raw topology data, but these approaches are intrinsically transductive and are unable to generalize to unseen data~\cite{DBLP:conf/nips/HamiltonYL17}. 

GNN-based link predictors, \eg SEAL~\cite{DBLP:conf/nips/ZhangC18}, inductively predict the edges, outperforming numerous non-parametric methods across diverse domains~\cite{DBLP:conf/nips/HamiltonYL17}.
To avoid computation, many GNNs~\cite{DBLP:conf/aaai/ShiHZHZZ24} adopt a local enclosing sub-graph centered on a pair of nodes to predict the edge between them.
But this ignores long-distance dependency, which is more frequently required in realistic scenarios~\cite{DBLP:conf/iclr/LiaoBTGUZ18}.

The exploitation of global information has proved a notable advantage of GNN-based methods, typically achieved through graph sampling~\cite{DBLP:conf/www/HuDWS20} or partitioning~\cite{DBLP:conf/kdd/ChiangLSLBH19}.
\textbf{(1) Graph sampling.} In GraphSAGE~\cite{DBLP:conf/nips/HamiltonYL17}, the context for the GNN learner is uniformly sampled across each order of neighbors, while GeniePath~\cite{DBLP:conf/aaai/LiuCLZLSQ19} employs a learned attention function to sample fixed-sized paths. 
Sampling approaches, however, may omit critical local or global information by neglecting useful nodes. 
\textbf{(2) Graph partitioning.} To mitigate the computational expense of learning global information, some methods divide non-sparse graphs into dense clusters to limit the context, as in ClusterGCN~\cite{DBLP:conf/kdd/ChiangLSLBH19}, or partition the graph to extract local and global information separately, as in GPNN~\cite{DBLP:conf/iclr/LiaoBTGUZ18}. 
Nevertheless, partition-based methods overlook the valuable global information of bridging structures between clusters. 

Instead of sampling or partitioning, we introduce virtual edges to compress global information for a lower computational cost, which parallels data augmentation techniques~\cite{DBLP:conf/aaai/0003LNW0S21}.
GAug~\cite{DBLP:conf/aaai/0003LNW0S21} adds or removes edges deterministically to generalize the model, while GCA~\cite{DBLP:conf/www/0001XYLWW21} generates multiple graph views with random noise to highlight the important connective structures.
MotifRGC~\cite{DBLP:conf/aaai/SunHWW0Y24} and LS-GCL~\cite{DBLP:journals/nn/YangLZDZ24} further leverage generative-contrastive learning to enrich node embeddings for link prediction.
PageRank Bandits (PRB)~\cite{NEURIPS2024_25ead0ef} frames link prediction as interactive network exploration via contextual bandits to capture long range dependencies from a different paradigm.
In contrast, our virtual edges compress the real long-distance dependencies, incorporating more relevant global information.

\subsection{SIR model}
The SIR model is originally used for epidemic trend analysis~\cite{DBLP:journals/amc/HarkoLM14}, but the SIR model and its derivative models are also frequently used in the field of social computing to analyze the information propagation process~\cite{DBLP:journals/socomp/WongJ22}, and maximize information diffusion~\cite{DBLP:journals/eswa/KumarMP23}.
For instance, with the adaptive differential evolution and particle swarm optimization, the SIR model could fit the basis of parameter changes over time to reflect the actual epidemic condition~\cite{DBLP:journals/socomp/WongJ22}.
GLSTM~\cite{DBLP:journals/eswa/KumarMP23} calculates node feature vectors and every node's individual influence under the susceptible-infected-recovered information diffusion model for the model to predict the probable influence of every node in the target network.
These kinds of research give insight into our method and show examples of how the SIR model can be combined with propagation-based analysis.


\section{Method overview}\label{sec:overview}

We aim to track all the relevant node information by a sub-graph whose shape, according to the proposed propagation model, is determined by the nodes defined in \therefs{the:H_out}{the:H_in} which can be approximated using \cororef{the:H_out_approx} and \eqaref{eq:noticed-set-approx} with proper parameters.
Next, \agoref{alg:extract_the_virtual_sub-graph} generates the corresponding compressed virtual sub-graph to train the model with reduced training cost, which is analyzed in \theref{the:ve_speedup}.
How the generated sub-graphs with different parameters affect the model effectiveness and how the virtual edges improve the model training efficiency are evaluated in the experiment section.
The overall \frameworkname framework, namely \framework, is shown in \figref{fig:the_framework}.
The following will introduce the notations and the problem we want to address.

\widefig{
\centerline{\includegraphics[width=1.0\textwidth]{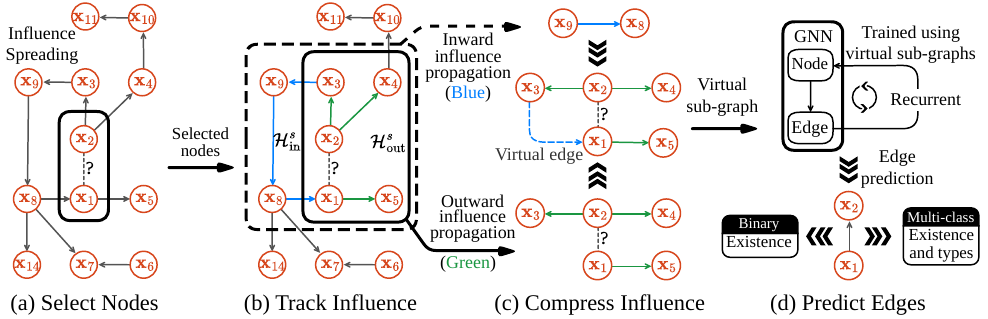}}
\caption{The main architecture of \framework.
\textbf{(a) Select Nodes}: We select a pair of nodes to predict the edge between them.
\textbf{(b) Track Influence}: We find a sub-graph containing the outward-propagated influence $\hout(s)$ and inward-propagated influence $\hin(s)$ via an extended propagation model; nodes of $\hout$ and $\hin$ are rendered in distinct colors, and border nodes are outlined (\secref{subsec:set-of-outward-influence-Hout},~\secref{subsec:set-of-inward-influence-Hin}).
\textbf{(c) Compress Influence}: Dashed arrows denote \textit{virtual} edges that replace paths of inward influence propagation; they are merged with the outward sub-graph to yield the compact \textit{virtual} sub-graph (\secref{subsec:generation-of-the-virtual-sub-graph}).
\textbf{(d) Predict Edges}: A recurrent GNN is trained on $\fsgraph$ (virtual sub-graph with features) and $\lsgraph$ (virtual sub-graph with labels) to predict edge existence and type (\secref{subsec:link-prediction-using-a-GNN}).
\label{fig:the_framework}
}
}

A directed graph is defined as $\graph = \abkt{\nodeset, \edgeset}$, where $\nodeset = \cbkt{\node(1), \dots, \nodei, \dots}$ is the set of nodes for each $\nodei \in \real[\realdim]$ indicates the $i$th node with the explicit (observable) attributes or (and) the implicit embedding, and $\edgeset = \cbkt{\edgeij: \adj(ij) \neq 0}$ is the set of edges for each $\edgeij \in \cbkt{1, \ldots, \numedgetype}$ that represents an edge from $\nodei$ to $\nodej$ with one of $\numedgetype$ edge types ($\numedgetype = 1$ for the homogeneous graphs without edge types).
$\adj$ is the adjacent matrix of $\graph$.
For $\nodei$ in a graph $\graph$, we denote by $\neighbor[\nborder](i)$ its set of neighbors of order $\nborder$ (\ie $\nborder$ ``hops'' away along the edge direction) and $\neighbor[0](i)$ is $\nodei$ itself.
$\pathsij$ represents the set of all the paths from node $\nodei$ to $\nodej$.

The task of link prediction is to complete or correct any missing or erroneous edges of $\graph$, given the corresponding features.
Focusing on a pair of nodes $\nodei$ and $\nodej$, the goal is to obtain $\featfunc{\cdot}$ which extracts relevant node information and $\lpfunc{\cdot}$ to predict the edges $\edgeij$:
\begin{equation}
    \label{eq:link_prediction}
\edgeij= \lpfunc{\featfunc{\nodei}, \featfunc{\nodej}},
\end{equation}

More notations of our method are shown in \tabref{tab:notations}.

\normaltab[t]{
\caption{The notations of our method.}
\label{tab:notations}
\scriptsize
\setlength{\parindent}{-3em}
\begin{tabular}{l l}
\toprule
Notation & Description \\
\midrule
$\graph = \abkt{\nodeset, \edgeset}$ & Directed graph including node set $\nodeset$ and edge sets $\edgeset$ \\
$\numedgetype$ & Number of edge types or relations \\
$\neighbor(s)$ & Set of neighbors of $\node(s)$ \\
$\hout(s)$ & Outward influence propagation node set of $\node(s)$ \\
$\hin(s)$ & Inward influence propagation node set of $\node(s)$ \\
$\noticedset(s)$ & Noticed node set of $\node(s)$ \\
$\border(s)$ & Border node set of $\node(s)$ \\
$\rbkt{\nodeset, \sigfd, \mInf(s) \rbkt{\cdot}}$ & Entire set of nodes, its $\sigma$-field, and the measure of any subset of the $\sigma$-field  \\
$\mS(s)\rbkt{\cdot}$, $\mI(s)\rbkt{\cdot}$, $\mR(s)\rbkt{\cdot}$ and $\mU(s)\rbkt{\cdot}$ & Measures of $\normal$, $\influential$, $\influenced$, and \noticed nodes, respectively \\
$\adj$ and $\diagmat$ & Graph adjacency matrix and diagonal degree matrix, respectively \\
$\infprob{ij}$ & Conditional activation probability \\
$\paths(ij)$ & All the possible paths from $\node(i)$ to $\node(j)$ \\
$\recovery(i)$ &  Duration for $\nodei$ converted from $\influential$ to $\influenced$ node \\
$\nborder$ and $\Q$&  Positive scalars to approximate $\mR[\tsp](s)\rbkt{\nodeset}$ and the relative size of $\hin(s)$ over $\hout(s)$, respectively  \\
$\fsgraph$ and $\lsgraph$ &  Sub-graphs with features and labels, respectively \\
$\recurrence$ &  Number of message passing iterations \\
$\loss$ &  Loss function to train the model \\
$\modelparam$ and $\netweight[\rbkt{\layers}]$ &  Model parameters and model weights of $\layers$-th layer, respectively \\
$\evals$ &  Latest $2 \lastepoch$ epochs of the evaluations during model training \\
$\evalthsh$ &  Threshold for early stop during model training \\
$\gsparsity$ & Average number of edges per node \\
$\realdim$ & Dimension of node features \\
$\ngbat$ & Number of sub-graphs in a mini-batch during model training \\
\bottomrule
\end{tabular}
}

\section{Tracking sub-graph of influence propagation}\label{sec:influence-propagation-theory}
All types of the propagated influence of a source node $\source$ will be held in a sub-graph, denoted as $\graph(s) \subset \graph$.
We treat $\graph(s)$ as features of $\source$ to predict edges, \ie, $\featfunc{\source} = \graph(s) = \abkt{\hall(s), \cbkt{\edgeij: \edgeij \in \edgeset; \nodei, \nodej \in \hall(s)}}$.
To find the $\featfunc{\source} = \graph(s)$ of $\source$ from $\graph$, we extend the SIR model~\cite{DBLP:journals/amc/HarkoLM14} for epidemic spreading to depict the influence propagation as a bi-directional process involving: (1)~the node set of outward influence propagation from $\source$, we denote by $\hout(s)$, see \secref{subsec:set-of-outward-influence-Hout}, and (2)~the node set of inward influence propagation (reversely responding the outward influence back to $\source$), denoted as $\hin(s)$, see \secref{subsec:set-of-inward-influence-Hin}.
Our objective is to find nodes in $\hall(s) = \hout(s) \cup \hin(s)$ from each $\source$.
Moreover, we do not distinguish the types of influence at this stage, and the relationship between types of influence and edges will be explored in \secref{sec:the-framework}.

\subsection{Node set of outward influence: \texorpdfstring{$\hout(s)$}{h-s-out}}\label{subsec:set-of-outward-influence-Hout}
Both the epidemic and the influence will propagate over the graph until reaching a stationary state.
We use $\hout(s)$ to include the nodes of outward influence propagation of $\source$ after the stationary state is reached.

According to the SIR model, three types of nodes depict the influence propagation process.
Given a state of the propagation from $\source$ at time $\ts$, nodes on the graph can be categorized as:
(1) $\normal$ nodes that can be influenced by the $\influential$ nodes.
(2) $\influential$ nodes that can propagate such influence to nearby $\normal$ neighbors for some time before transforming to the $\influenced$ nodes ($\source$ is the first $\influential$ node at the beginning of influence propagation).
(3) $\influenced$ nodes that cannot propagate influence to nearby $\normal$ nodes anymore.
Thus, all the $\influential$ and $\influenced$ nodes are those that have been influenced.

Formally, we consider a measurable space $\rbkt{\nodeset, \sigfd, \mInf(s) \rbkt{\cdot}}$ where $\sigfd$ is the $\sigma$-$\text{field}$ on $\nodeset$ and $\mInf(s)\rbkt{\cdot}$ is a measure on any subset of $\sigfd$ for the number of nodes.
Besides, finite measures $\mS(s)\rbkt{\cdot}$, $\mI(s)\rbkt{\cdot}$, and $\mR(s)\rbkt{\cdot}$, are the measures for the number of $\normal$, $\influential$, and $\influenced$ nodes respectively satisfying $\mInf[\ts](s)\rbkt{\nodeset} = \mS[\ts](s)\rbkt{\nodeset} + \mI[\ts](s)\rbkt{\nodeset} + \mR[\ts](s)\rbkt{\nodeset}$ at time $\ts$.
Then, by considering topology information $\adj(ij) \in \cbkt{0,1}$ from the adjacency matrix and conditional activation probability $\infprob{ij}$, which is the probability of $\nodej$ is influenced by $\nodei$ given the edge type $\edgeij$, we define our graph SIR model in \defref{the:Graph_SIR_Model}.
\defi[Graph SIR Model]{
    \label{the:Graph_SIR_Model}
    For any node $\source$, given measures $\mS(s)\rbkt{\cdot}$, $\mI(s)\rbkt{\cdot}$ and $\mR(s)\rbkt{\cdot}$ on $(\nodeset, \sigfd, \mInf(s)\rbkt{\cdot})$, the SIR model on graph $\graph$ over time $\ts \in \eqcbkt{0, 1, \ldots, \infty}$ is represented as:
    \begin{equation}
        \eqlcbkt{
            \begin{aligned}
                \displaystyle \frac{\dif}{\dif \ts} \mS[\ts](s)\rbkt{\nodeset} & = - \mathop{\sum}_{i \in \nodeset}\mathop{\sum}_{j \in \nodeset} \adj(ij) \infprob{ij} \mI[\ts](s)(i) \mS[\ts](s)(j) &  & \\
                \displaystyle \frac{\dif}{\dif \ts} \mI[\ts](s)\rbkt{\nodeset} & = - \frac{\dif}{\dif \ts} \mS[\ts](s)\rbkt{\nodeset} - \frac{\dif}{\dif \ts} \mR[\ts](s)\rbkt{\nodeset}              &  & \\
                \displaystyle \frac{\dif}{\dif \ts} \mR[\ts](s)\rbkt{\nodeset} & = \mathop{\sum}_{i \in \nodeset} \frac{1}{\recovery(i)} \mI[\ts](s)(i) \period                                       &  &
            \end{aligned}
        }
        \label{eq:SIR_trend_t}
    \end{equation}
    where $\recovery(i)$ is the duration for $\nodei$ from $\influential$ to $\influenced$ node, and $\infprob{ij}$ is the conditional activation probability for any $\nodei, \nodej \in \nodeset$ satisfying $\sum_{\nodej \in \neighbor[1](i)} \infprob{ij} = 1$.
}
\noindent In the graph SIR model, the propagation process, akin to an information cascade reaching saturation in social networks, will reach a stationary state at time $\tsp > 0$.
Specifically, $\mI[\ts](s)\rbkt{\nodeset}=0$ when $\ts \geq \tsp$, and $\mInf[\tsp](s)\rbkt{\nodeset} = \mS[\tsp](s)\rbkt{\nodeset} + \mR[\tsp](s)\rbkt{\nodeset}$, which means that the $\influential$ nodes will disappear after $\tsp$, leaving only the $\normal$ and $\influenced$ nodes.
In this stationary state, which starts at $\tsp$, the $\influenced$ nodes will be the set of $\hout(s)$, and $\abs{\hout(s)} = \mR[\tsp](s) \rbkt{\nodeset}$, since the set of the $\influenced$ nodes at $\tsp$ covers all possible paths for the outward influence propagation outbreaks from $\source$.
The $\influenced$ nodes include nodes in $\neighbor[1](s)$ and $\paths(sE)$, where $\node(E) \neq \source$ receives the influence of $\source$ and terminates the propagation.
\theref{the:H_out} shows the set of $\hout(s)$.
\theo{
    \label{the:H_out}
    Given $\mR[\tsp](s)\rbkt{\nodeset}$, and all possible paths $\paths(si)$ between $\source$ and $\nodei$, the most likely set of $\hout(s)$ of the source $\source$ will be
    \begin{equation}
        \label{eq:H_out}
        \hout(s) \stackrel{\prob}{\longrightarrow} \eqcbkt{ \nodei:
            \sum_{\singlepath \in \paths(si)} \prod_{\edge(ab) \in \singlepath} \infprob{ab}
            \ge \frac{1}{\mR[\tsp](s)\rbkt{\nodeset}} } \period
    \end{equation}
}

\theref{the:H_out} selects the number of nodes equal to the value of $\mR[\tsp](s)\rbkt{\nodeset}$, with the highest probability of receiving the outward influence from the source, as $\influenced$ nodes to obtain $\hout(s)$. Detailed proof of \theref{the:H_out} can be found in \appref{sec:proof-of-theorem-h-out}.

According to \theref{the:H_out}, $\mR[\tsp](s)\rbkt{\nodeset}$ is the key to obtain $\hout(s)$.
Actually, $\infprob{ij}$ is nearly uniform and independent of $\edgeij$ in many real networks~\cite{DBLP:conf/nips/ZhangC18}, and $\recovery(i)$ can be chosen as a constant based on the application.
Considering the graph topology $\adj(ij)$, the influence of $\source$ has been propagated only by existing edges to its neighbors $\neighbor(s)$, since edges are produced by the influence.
Thus, we give the following approximation for $\mR[\tsp](s)\rbkt{\nodeset}$ to obtain $\hout(s)$ in \cororef{the:H_out_approx}.

\zcoro{
    \label{the:H_out_approx}
    Given $\nodei$, $\nodej \in \nodeset$, let $\infprob{ij} = \uniprob$ be uniformly distributed, $\recovery(i) = \hat{\recovery}$ be a constant, there will be a proper $\nborder$ to approximate $\mR[\tsp](s)\rbkt{\nodeset}$:
    \begin{align}
        \label{eq:n_hat}
         & \displaystyle \nborder = \argmin_{\nborder[\prime]}
        \eqabs{
        \sum_{k=0}^{\nborder[\prime]} \eqabs{ \neighbor[k](s) } - \mR[\tsp](s)\rbkt{\nodeset}
        } \nonumber                                                                                                                                                                                                            \\
         & \displaystyle \sdtd \exp (- \hat{\recovery} \cdot \uniprob \cdot \eqabs{ \neighbor[1](s) } \cdot \frac{\mR[\tsp](s)\rbkt{\nodeset}}{\eqabs{\nodeset}}) = 1 - \frac{\mR[\tsp](s)\rbkt{\nodeset}}{\eqabs{\nodeset}} ,
    \end{align}
    which satisfies $\mR[\tsp](s)\rbkt{\nodeset} \approx \mR[\tsp](s)(\cup_{k=0}^{\nborder} \neighbor[k](s))$. Thus, $\hout(s) \approx \cup_{k=0}^{\nborder} \neighbor[k](s)$.
}

\cororef{the:H_out_approx} uses the local neighbor nodes of the source node to approximate the nodes in $\hout(s)$.
\cororef{the:H_out_approx} also explains why methods that employ information from a local range of neighbors exhibit satisfactory performance in certain real-world datasets, as these methods typically leverage the outward influence of the source.
Detailed proof of \cororef{the:H_out_approx} can be found in \appref{sec:proof-of-h-out-approx}.
To consider the node influence completely, we next discuss the propagation of inward influence.

\subsection{Node set of inward influence: \texorpdfstring{$\hin(s)$}{h-s-in}}\label{subsec:set-of-inward-influence-Hin}
After the outward influence propagation, some influences of $\source$ continuously propagate outside $\hout(s)$ as responses before going back to $\source$, which still need to be tracked.
In this section, we uncover the nodes of inward influence, covering the nodes set $\hin(s)$, which reversely respond to the outward influence from $\source$.

The paths of response will be only within a range of nodes that have noticed $\source$.
In particular, we extend the SIR model by introducing the \noticed nodes to include the nodes that are on the paths of response to the influence of the source node.
For example, in a social network, \noticed nodes are those aware of $\source$'s influence yet not directly altered by it, among whom some eventually give response to $\source$, forming $\hin(s)$.

We assume that the number of the \noticed nodes is positively correlated with the number of the nodes influenced by $\source$, and we define the measure of the \noticed nodes in \defref{the:noticed-node}.
\defi[Measure of the Noticed Nodes]{
    \label{the:noticed-node}
    Given a positive term $\Qt < \infty$, let $\mU[\ts](s) \rbkt{\cdot}$ be a measure for the number of \noticed nodes at $\ts \in \eqcbkt{0, 1, \ldots, \infty}$ satisfying $\mU[\ts](s)\rbkt{\cdot} = \Qt \cdot \rbkt{\mI[\ts](s) \rbkt{\cdot} + \mR[\ts](s) \rbkt{\cdot}}$, and then we have
    \begin{equation}
        \displaystyle \frac{\dif}{\dif \ts} \mU[\ts](s) \rbkt{\nodeset} = \sum_{i \in \nodeset} \eqrbkt{1 + \frac{1}{\recovery(i)}} \cdot \Qt \cdot \mI[\ts](s) \rbkt{i} \period
        \label{eq:U-trend-t}
    \end{equation}
}

All the paths of response can be covered by the set of \noticed nodes, denoted as $\noticedset(s)$.
By searching the paths of response in $\noticedset(s)$, we could obtain $\hin(s)$ in \theref{the:H_in}.
\theo{
    \label{the:H_in}
    Given $\sum_{\nodej \in \neighbor[1](\ast)} \infprob{\ast j} = 1$, $\paths(si)$, and $\mR[\tsp](s)\rbkt{\nodeset}$, let $\mU[\tsp](s)\rbkt{\nodeset} = \Qt \cdot (\mI[\tsp](s)\rbkt{\nodeset} + \mR[\tsp](s)\rbkt{\nodeset}) = \Qt \cdot \mR[\tsp](s)\rbkt{\nodeset}$, and $\singlepath$ is a single path from $\paths(si)$, $\hin(s)$ will be
    \begin{equation}
        \displaystyle \hin(s) = \bigcup_{j \in \border(s), i \in \noticedset(s)} \eqcbkt{\node(k): \node(k) \in \paths(ji) ; \node(k) \ne \nodej ; \adj(is) \ne 0} \period
    \end{equation}
    where $\noticedset(s)$ is the set of the \noticed nodes
    \begin{equation}
        \label{eq:noticed-set}
        \noticedset(s) \!\!\stackrel{\prob}{\longrightarrow}\!\! \eqcbkt{ \nodei:
            \frac{\frac{1}{\eqrbkt{\Qt + 1}}}{\mR[\tsp](s)\rbkt{\nodeset}} \le \!\!\!\!
            \sum_{\singlepath \in \paths(si)} \! \prod_{\edge(ab) \in \singlepath} \infprob{ab}
            \!<\! \frac{1}{\mR[\tsp](s)\rbkt{\nodeset}}
        } \period
    \end{equation}
}

$\border(s)$ in \theref{the:H_in} is defined as the last order neighbors of $\source$ in $\hout(s)$, which split the $\hall(s)$ into $\hout(s)$ and $\hin(s)$, given in \defref{the:border_node}.
\defi[Border Set]{
    \label{the:border_node}
    Let $\nborder$ satisfy $\hout(s) \approx \cup_{k=0}^{\nborder} \neighbor[k](s)$, and $\paths(sj)$ be all the paths from $\source$ to $\nodej$ in $\nodeset$, the border set of $\source$ is
        {\small \begin{equation}
                \border(s) = \eqcbkt{
                    \nodei: \min_{\singlepath \in \paths(si)} \eqabs{\singlepath} = \nborder
                } \period
                \label{eq:border_node}
            \end{equation}}
}

\theref{the:H_in} determines the \noticed node set according to the probability of receiving the inward influence of the source, and finds nodes in the paths back to the source among \noticed nodes to obtain $\hin(s)$. 
In \theref{the:H_in}, $\node(k) \in \paths(ji)$ are the nodes of all possible paths from $\nodej$ back to $\nodei$.
Using \therefs{the:H_out}{the:H_in}, the source node's influence can be tracked. Detailed proof of \theref{the:H_in} can be found in \appref{sec:proof-of-theorem-Hin}.

Using the similar approximation method in \cororef{the:H_out_approx}, we can approximate $\noticedset(s)$ by
\begin{equation}
    \label{eq:noticed-set-approx}
    \displaystyle \noticedset(s) \approx \bigcup_{k=\nborder+1}^{\nborder \times \eqrbkt{\Q + 1}} \neighbor[k](s) ,
\end{equation}
where $\Q$ is a positive scalar to approximate $\Qt$ that indicates the co-relationship between the node set size of outward and inward influence propagation.
Experiments show a proper scalar can be good enough to obtain the $\noticedset(s)$.

Consequently, by the approximation of \therefs{the:H_out}{the:H_in} using \cororef{the:H_out_approx} and \eqaref{eq:noticed-set-approx} with the proper $\nborder$ and $\Q$, we filter irrelevant noises and obtain the relevant information from nodes in $\hall(s) = \hout(s) \cup \hin(s)$ to track the influence of $\source$ covered by $\graph(s)$ for link prediction.

\section{Sub-graph of influence propagation for link prediction}\label{sec:the-framework}
In this section, we present our \framework as $\lpfunc{\cdot}$ to predict the types of the edge with a lower computational cost, by using the virtual sub-graph $\vsgraph(ij)$ obtained from $\graph(ij) = \graph(i) \cup \graph(j)$.
To this end, we first replace the paths of inward influence propagation by virtual edges to obtain virtual sub-graphs to reduce the computational costs in \secref{subsec:generation-of-the-virtual-sub-graph}.
Subsequently, we train a recurrent GNN taking virtual sub-graphs as input to predict edge types in \secref{subsec:link-prediction-using-a-GNN}.
Efficiency analysis of using virtual edges is given in \secref{subsec:analysis}.

\subsection{Virtual sub-graph generation}\label{subsec:generation-of-the-virtual-sub-graph}
Usually, $\hin(ij) = \hin(i) \cup \hin(j)$ from $\graph(ij)$ covers a large range and leads to a high computational cost.
Thus, we introduce the \emph{virtual} edges to simulate the paths of influence propagation through $\hin(ij)$ to reduce the computational cost.

The concept of the virtual edge is derived from the Mean-Field Theory (MFT)~\cite{pastor2015epidemic}, which uses the average effect to replace complex group effects in a simplified and computationally efficient manner.
According to MFT, the expectation of inward influence propagates from nodes in $\border(i)$ back to $\nodei$ over $\hin(i)$, \ie $\sum_{\singlepath \in \paths(bi)} \prod_{\edge(\ast) \in \singlepath} \E \eqbbkt{\adj(\ast) \cdot \infprob{\ast}}$ ($\node(b) \in \border(i)$), could be considered as a group effect of $\E \eqbbkt{\infprob{bi}}$.
A virtual edge $\edge(bi)$, which virtually acts as these paths of inward influence propagation, can replace such a group effect.
We focus on the topological reachability from $\border(i)$ back to $\nodei$ over $\hin(i)$ instead of nodes, since the nodes over $\hin(i)$ usually bring much noise due to their less relevant information compared to the source.
Besides, we only replace paths over $\hin(i)$ rather than $\hall$, since the paths over $\hin(i)$ carry more relevant information than nodes far from the source.
By using virtual edges, the $\hall(ij)$ can be enclosed by $\hout(ij)$, and $\graph(ij)$ will be shrunk to the sub-graph of outward influence propagation augmented with virtual edges, \ie, the virtual sub-graph $\vsgraph(ij)$.
Utilizing $\vsgraph(ij)$ will not dramatically increase the computational cost, while the relevant global information of $\nodei$ and $\nodej$ is also reserved.

\alg[t]{Virtual sub-graph generation}{  
    \label{alg:extract_the_virtual_sub-graph}
    \Input{$\graph$, $\adj$, $\nodei$ (the source) and $\node(j)$, $\nborder$, $\Q$.}
    \Output{The pair of virtual sub-graphs $\fsgraph(ij)$ and $\lsgraph(ij)$.}
    Find $\hout(i)$ and $\hout(j)$ by using $\graph$ according to \cororef{the:H_out_approx}.\;
    Obtain $\vsgraph(ij) = \hout(i) \cup \hout(j)$.\;
    Find $\border(ij) = \border(i) \cup \border(j)$ in $\hout(i) \cup \hout(j)$ according to \eqaref{eq:border_node}.\;
    Obtain $\noticedset(ij) = \noticedset(i) \cup \noticedset(j)$ using $\Q$ according to \eqaref{eq:noticed-set-approx}.\;
    \ForEach{$\node(b) \in \border(ij)$}{
        Add $\node(b)$ to an empty set $\candset$ to store candidate nodes.\;
        Add $\neighbor[1](b)$ to an empty set $\visitedset$ to store visited nodes.\;
        \While{$\abs{\visitedset} < \abs{ \noticedset(ij) }$}{
            Add $\node(l) \notin \candset$ with $\edgefeat(li) = \underset{\node(k) \in \visitedset}{\min} \rbkt{\edgefeat(ki)}$ to $\candset$.\;
            Add $\neighbor[1](l) \cap \noticedset(ij) $ to $\visitedset$.\;
            \If{$\visitedset$ contains $\node(m)$ with $\adj(mi) = 1$}{
                Add $\edge(bi)$ to $\vsgraph(ij)$ and search for the next $\node(b)$.\;
            }
        }
    }
    Duplicate $\vsgraph(ij)$ into $\fsgraph(ij)$ and $\lsgraph(ij)$.\;
    Assign the feature to $\edge(ab)\in \fsgraph(ij)$ with $\edgefeat(ab)$.\;
    Assign the label to $\node(a) \in \lsgraph(ij)$, \text{True} if $\node(a) \in \cbkt{\nodei, \nodej}$ otherwise \text{False}.\;
    Return the pair of virtual sub-graphs $\fsgraph(ij)$ and $\lsgraph(ij)$.\;
}

\agoref{alg:extract_the_virtual_sub-graph} converts $\graph(ij)$ into $\vsgraph(ij)$ by virtual edge extraction and adds features/labels to $\vsgraph(ij)$ to train the edge predictor.
We assume that $\nodei$ and $\nodej$ are the source and destination of the edge to predict.
\agoref{alg:extract_the_virtual_sub-graph} first finds $\border(ij)$ of $\hout(ij)$ and $\noticedset(ij)$.
Then for $\node(b)$ in $\border(ij)$, we add a virtual edge from $\node(b)$ to $\nodei$ in $\vsgraph(ij)$ only if a path exists in $\noticedset(ij)$ connecting these two, enabling us to obtain the structure of $\vsgraph(ij)$.
During the path searching, we use node features to guide the search for paths.
The process starts by taking any $\node(b) \in \border(ij)$ as the first candidate node of $\paths(bi)$.
The greedy search selects the first-order neighbor of the candidate node nearest to $\nodei$ as the next candidate, repeating until $\paths(bi)$ is confirmed or refuted; since at most $\abs{\noticedset(ij)}$ nodes are traversed and each step requires an $\bigO\rbkt{\abs{\noticedset(ij)}}$ nearest-neighbor scan, the worst-case time complexity of \agoref{alg:extract_the_virtual_sub-graph} is $\bigO\rbkt{\abs{\border(ij)} \cdot \abs{\noticedset(ij)}[2]}$, reducible to $\bigO\rbkt{\abs{\border(ij)} \cdot \abs{\noticedset(ij)} \log \abs{\noticedset(ij)}}$ via a min-heap.
Once the structure of $\vsgraph(ij)$ has been obtained, two replicas of it will be made to fill features and labels respectively as $\fsgraph(ij)$ and $\lsgraph(ij)$ for the later model training.
Note that the $\fsgraph(ij)$ contains the information of all types of propagated influence and can be utilized by the model.
In $\fsgraph(ij)$, the node feature $\nodei$ is its implicit or (and) explicit features, and the edge feature $\edgefeat(ij)$ is a Euclidean or cosine distance between $\nodei$ and $\nodej$.
In $\lsgraph(ij)$, the edge labels are the existence for binary prediction or types for multi-class prediction.
We also add node labels to $\lsgraph(ij)$, which masks the selected nodes $\nodei$ and $\nodej$, to check whether the node category can affect the performance.
Notably, the type of the virtual edge is a special type that is different from all the other types.
\figref{fig:feat_lab_virtual_sub_graph} gives an example to show $\fsgraph$ and $\lsgraph$ for $\vsgraph$ in \figref{fig:the_framework}(c).

\normalfig[tb]{
\subfloat[$\fsgraph$ ($\vsgraph$ with features)]{
\includegraphics[scale=0.75]{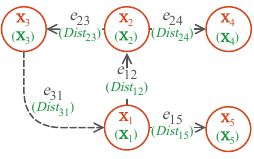}
\label{fig:virtual_edge_input}
}
\subfloat[$\lsgraph$ ($\vsgraph$ with labels)]{
\includegraphics[scale=0.75]{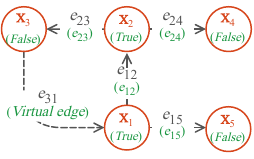}
\label{fig:virtual_edge_target}
}
\caption{Featured and labeled virtual sub-graphs of the virtual sub-graph in \figref{fig:the_framework}(c) with $\nborder=1$ for multi-class prediction. Dashed lines are virtual edges. Features in $\fsgraph$ and labels in $\lsgraph$ are in green.}
\label{fig:feat_lab_virtual_sub_graph}
}

\subsection{The propagated-influence-based link prediction}\label{subsec:link-prediction-using-a-GNN}
Once we have $\fsgraph$ and $\lsgraph$, we train the recurrent GNN~\cite{DBLP:journals/corr/abs-1806-01261} to map the features describing all types of influence in $\fsgraph$ to their corresponding edge types in $\lsgraph$ via message passing.

\parahead{Recurrent GNN.} A recurrent GNN performs two steps recurrently in each layer: (1)~edges are updated using messages from connected nodes, and (2)~nodes are updated using the messages aggregated from their edges.
This process can be depicted in \eqaref{eq:gnn}.
\begin{equation}
    \label{eq:gnn}
    \nodefeat[\rbkt{\layers + 1}] = \sigma(\diagmat[-1] \adj \nodefeat[\rbkt{\layers}] \netweight[\rbkt{\layers}]) \rbkt{0 < \layers < \recurrence}
\end{equation}
where $\sigma(\cdot)$ is the update function, the recurrence $\recurrence$ indicates the number of message passing iterations, $\diagmat$ is the diagonal degree matrix, and $\diagmat[-1] \adj$ is the normalized adjacency matrix.
$\nodefeat[\rbkt{0}]$ and $\nodefeat[(\recurrence)]$ are the input feature and final representation of nodes, respectively.
A decoder will use the final node representation to generate edge embeddings and output the types of every edge in the virtual sub-graph.
Different $\recurrence$ affects the learning process, which can be found in the experiments.

\parahead{Training.} We use the supervised learning scheme to train the recurrent GNN to predict the edges using $\fsgraph(ij)$ and $\lsgraph(ij)$ obtained by the extended propagation model and \agoref{alg:extract_the_virtual_sub-graph}.
The recurrent-GNN fulfills the message passing guided by the graph structure of influence propagation from the $\fsgraph(ij)$, and subsequently outputs the edge existence and type, \ie $\edgeij$.
The training process is more effective and efficient given the light but comprehensive global information with long-distance dependencies stored in the virtual sub-graph.
To show how edges and nodes information of the virtual sub-graphs contribute to the final prediction results, different losses derived from a cross-entropy loss $\loss[CE]$~\cite{DBLP:journals/corr/abs-1806-01261} are designed concerning model parameters $\modelparam$: (1) $\loss(0)$ considering labels of both nodes and edges in $\vsgraph(ij)$ in \eqaref{eq:loss_nodes_edges}; (2) $\loss(1)$ only considering labels of edges in $\vsgraph(ij)$ in \eqaref{eq:loss_edges}; (3) $\loss(2)$ only considering the label of $\edgeij$ in \eqaref{eq:loss_edges_ij} which is the original loss for link prediction.
The objective of the training process is to minimize the loss.
{\small
\begin{equation}
    \label{eq:loss_nodes_edges}
    \loss(0) \eqrbkt{\modelparam} = \sum_{\edgeij \in \edgeset} \eqrbkt{
        \sum_{\node(a) \in \vsgraph(ij)} \loss[CE] \eqrbkt{\node(a), \node[\modelparam](a)}
        +
        \sum_{\edge(ab) \in \vsgraph(ij)} \loss[CE] \eqrbkt{\edge(ab), \edge[\modelparam](ab)}
    }
\end{equation}
\begin{equation}
    \label{eq:loss_edges}
    \loss(1) \eqrbkt{\modelparam} = \sum_{\edgeij \in \edgeset}
    \sum_{\edge(ab) \in \vsgraph(ij)} \loss[CE] \eqrbkt{\edge(ab), \edge[\modelparam](ab)}
\end{equation}
\begin{equation}
    \label{eq:loss_edges_ij}
    \loss(2) \eqrbkt{\modelparam} = \sum_{\edgeij \in \edgeset}
    \loss[CE] \eqrbkt{\edge(ij), \edge[\modelparam](ij)}
\end{equation}
}

Besides, we also use \eqaref{eq:termination} to detect the convergence of the model for an early stop.
As such, the last $2 \lastepoch$ ($\lastepoch \in \cbkt{0, 1, \ldots}$) epochs of evaluations (\ie accuracy), denoted as $\evals$ ($\eqabs{\evals} \ge 2 \lastepoch$), will be used to stop the training by a designated threshold $\evalthsh$ in \eqaref{eq:termination}.
{\small
\begin{equation}
    \frac{1}{\lastepoch} \sum_{i= \eqabs{\evals} - 2 \lastepoch}^{\eqabs{\evals} - \lastepoch} \evals(i) - \frac{1}{\lastepoch} \sum_{j = \eqabs{\evals} - \lastepoch}^{\eqabs{\evals}} \evals(j) \le \evalthsh.
    \label{eq:termination}
\end{equation}
}

\parahead{Inference.} The trained model will process the input sub-graph $\fsgraph(ij)$, from a test graph $\graph[\prime]$, with explicit or (and) implicit features of a pair of nodes $\nodei$ and $\nodej$, to make the data-driven prediction for edge existence and type.
This can be written as $\lpfunc{\featfunc{\nodei}, \featfunc{\nodej}} = GNN \rbkt{\fsgraph(ij)} = \edgeij$.

\subsection{Analysis}\label{subsec:analysis}
Next, we show the complexity advantage of GNN training after adopting our \emph{virtual} sub-graphs.
We first give the training time complexity of common GNNs in \lemref{the:complexity_of_GNNs} for the sub-graphs of $\nborder$ order neighbors, whose proof can be found in \appref{sec:proof-of-lemma-gnn-comlexity}.
\lemm[Complexity of GNNs]{
    \label{the:complexity_of_GNNs}
    Given the graph sparsity $\gsparsity = \frac{\abs{\edgeset}}{\abs{\nodeset}}$ (the average number of edges per node), the training time complexity of a GNN is $\bigO \rbkt{ \ngbat \cdot \sum_{i=0}^{\nborder} \gsparsity[i] \cdot \realdim[2] \cdot \recurrence}$ where $\recurrence$ is the number of GNN layers or recurrence, $\realdim$ is the dimension of node features, $\nborder$ is the order of neighbors, and $\ngbat$ is the number of sub-graphs in a mini-batch.
}

Next, we discuss the speed-up of GNN training with virtual sub-graphs.
Suppose we take as input the nodes within $\nborder \cdot \rbkt{\Q + 1}$ order of neighbors for a GNN, the time complexity will be $\bigO \rbkt{ \ngbat \cdot \sum_{i=0}^{\nborder \cdot \rbkt{\Q + 1}} \gsparsity[i] \cdot \realdim[2] \cdot \recurrence}$.
If we use virtual edges to replace information far from $\nborder$ order of neighbors, the time complexity will drop to $\bigO \rbkt{\ngbat \cdot \sum_{i=0}^{\nborder} \gsparsity[i] \cdot \realdim[2] \cdot \recurrence}$.
Consequently, according to \lemref{the:complexity_of_GNNs}, we give the speed improvement after adopting the virtual sub-graph by \theref{the:ve_speedup}.
The proof of \theref{the:ve_speedup} can be found in \appref{sec:proof-of-speedup}.
\theo{
    \label{the:ve_speedup}
    By using the virtual sub-graph, the speed of GNN training is $\frac{\sum_{i=0}^{\nborder \cdot \rbkt{\Q + 1}} \gsparsity[i]}{\sum_{i=0}^{\nborder} \gsparsity[i]}$ times faster.
}

Above all, adopting the virtual sub-graph will considerably reduce the computational complexity of training a GNN model.

\section{Experiments}\label{sec:experiments}
We evaluate the performance of our \framework for multi-class and binary prediction in \secref{subsec:multi-class-prediction} and \secref{subsec:binary-prediction}, respectively.
Next, we conduct a rigorous analysis and ablation study in \secref{subsec:ablation-study} to show the efficiency and effectiveness improvement by adopting the virtual edges, the effect of inward influence, impacts of different losses, parameter analysis, and visualization of model training.
All experiments are conducted on a server with a 32-core CPU, 96~GB of RAM, and a GeForce RTX 2080 Ti GPU.

\parahead{Datasets.}\label{subsec:datasets}
The experiments are conducted on popular benchmarks and a real-world dataset.
(1) \ppi~\cite{DBLP:conf/nips/HamiltonYL17} is a protein-protein interaction database.
We use the first 3k nodes for the experiment.
(2) \dblp~\cite{DBLP:conf/icdm/YangL12} is a co-authorship network where two authors are connected if they publish at least one paper together.
(3) \blogcatalog~\cite{DBLP:journals/pvldb/YangSX0LB20} is a social network that stores the relationship between bloggers.
(4) \twitter~\cite{DBLP:conf/nips/McAuleyL12} consists of many ego networks.
An ego network contains an ego user and users to whom this ego user is directly connected.
Given that ego networks are largely independent and exhibit minimal interconnection, we focus on using a single ego network in our experiments.
(5) \wiki~\footnote{\url{https://dumps.wikimedia.org/wikidatawiki/entities/}} is the knowledge graph for Wikipedia actively maintained by the Wikimedia Foundation. 
Different subsets of entities are selected for different tasks, and all edges whose target entities lie outside the chosen subset are removed accordingly.
The resulting isolated nodes are retained for the negative sample generation, thereby increasing the difficulty of accurately predicting relationships among nodes.

\normaltab{
\caption{Statistics of datasets. $\abs{\nodeset}$ and $\abs{\edgeset}$ are the number of nodes and edges we used. $\numedgetype$ is the number of edge types or relations. $\nodei$ (with $\realdim$ dimensions) is either the pre-trained embeddings or observable features for \framework.}
\label{tab:statistics_of_datasets}
\begin{tabular}{l rrrrr}
\toprule
Dataset              & $\abs{\nodeset}$ & $\abs{\edgeset}$ & $\numedgetype$ & $\nodei$ for \framework      & $\realdim$ \\
\midrule
(1) \ppi             & 3000             & 5273             & 2              & N2V                          & 64         \\
(2) \dblp            & 317080           & 1049866          & 2              & N2V                          & 500        \\
(3) \blogcatalog     & 5196             & 343486           & 2              & N2V                          & 500        \\
(4) \twitter         & 179              & 6741             & 2              & In Dataset                   & 500        \\
(5) \wiki (1k/5k/10k)& 1k/5k/10k        & 369/1651/8248    & 260/380/420    & W2V                          & 500        \\
\bottomrule
\end{tabular}
}

\parahead{Data preparation.}\label{subsec:data-preparation}
The statistics of the used datasets are in \tabref{tab:statistics_of_datasets}.
Different datasets use different node features for our \framework.
For datasets (1)--(3), we use Node2vec~\cite{DBLP:conf/kdd/GroverL16} (N2V) to obtain pre-trained embeddings as node features.
Dataset (4) contains observed browser data where the top 500 frequently appearing attributes are filtered for use.
Dataset (5) uses \textit{wikipedia2vec}~\cite{DBLP:conf/conll/YamadaS0T16} (W2V) as the node feature.

Following the practice of existing works~\cite{DBLP:conf/aaai/ChenPHS18,DBLP:conf/nips/ZhangC18}, we equally split datasets (1)--(5) into two parts for training and testing, respectively.
The ratio of positive and negative samples was set to 1:1 for balanced label distribution and fairly binary prediction.
There are no negative samples for multi-class prediction.
The order of the training set is randomized across mini-batches to avoid any training bias.

\parahead{Comparison methods.}\label{subsec:methods-for-comparison}
To ensure a rigorous and effective comparison, we comprehensively benchmark both classical and recent approaches under a unified experimental setup.
More precisely, we compare our \framework with similarity metrics including AA~\cite{DBLP:journals/socnet/AdamicA03}, RA~\cite{zhou2009predicting}, latent feature methods including Node2vec~\cite{DBLP:conf/kdd/GroverL16} (N2V), Struc2vec~\cite{DBLP:conf/kdd/RibeiroSF17} (S2V), HARP~\cite{DBLP:conf/aaai/ChenPHS18} (based on Node2vec), Ripple2Vec~\cite{Luo2022ripple2vecNE} (R2V), and GNN-based methods including SEAL~\cite{DBLP:conf/nips/ZhangC18} (based on Node2vec, and $\nborder$ auto-selection is enabled), RGCN~\cite{DBLP:conf/esws/SchlichtkrullKB18}, HGT~\cite{DBLP:conf/www/HuDWS20}, GAug~\cite{DBLP:conf/aaai/0003LNW0S21}, GCA~\cite{DBLP:conf/www/0001XYLWW21}, BUDDY~\cite{chamberlain2022graph}, NCN~\cite{wang2024neural}, NCN-diff~\cite{wang2024neural}, NCNC~\cite{wang2024neural}, and Refined-GAE~\cite{ma2025reconsidering}.
Similarity and stochastic metrics indicate the performance of the traditional link prediction.
The comparison with latent feature methods evaluates the global feature extraction ability of our method.
We also include GNN-based methods, emphasis on different functionalities, to compare the learning ability among deep models.
See \appref{sec:baselines} for more baseline details.

\normalfig{
\subfloat[$\fsgraph$ ($\vsgraph$ with features)]{
\includegraphics[scale=0.75]{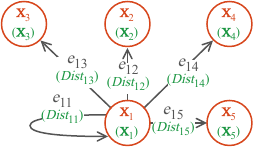}
\label{fig:center_beam_input}
}
\subfloat[$\lsgraph$ ($\vsgraph$ with labels)]{
\includegraphics[scale=0.75]{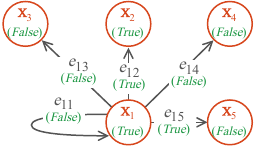}
\label{fig:center_beam_target}
}
\caption{Examples of the center beam (\textit{CB}) sub-graphs of the input graph in \figref{fig:the_framework}(a) for binary prediction.}
\label{fig:center_beam}
}

For ablation studies, we implement two variants of \framework using different approaches to generate sub-graphs.
The first variant uses the center-beam sub-graph structure (\textit{CB}), shown in \figref{fig:center_beam}, which only considers the embedding distance between $\source$ and nodes in $\hout$, neglecting neighbor structure.
\textit{CB} tries to learn a distance threshold to predict the edge, similar to methods that utilize features in a latent space.
\textit{CB} also demonstrates whether embeddings can restore graph structure.
Another variant uses the plain sub-graph structure (\textit{PL}), which covers $\hout$ with the original graph structure and node features, similar to SEAL~\cite{DBLP:conf/nips/ZhangC18}.
For comparison, our \framework utilizes nodes from both \therefs{the:H_out}{the:H_in}.

\parahead{Evaluation metrics.}\label{subsec:evaluation-metrics}
To comprehensively demonstrate the performance, we adopt accuracy, ROC-AUC, and average precision (AP) for binary prediction, and accuracy (Acc) and macro-F1 for multi-class prediction.
Except for efficiency experiments, all results are obtained using early stop defined by \eqaref{eq:termination} with $\lastepoch = 10$ and $\evalthsh = 1e-4$.
When compared with other methods, all results are averaged over three runs.

\widetab[tb]{
    \caption{Comparison of multi-class and binary link prediction. -- indicates the task is not supported by the method. All baselines are re-evaluated and fine-tuned under our unified experimental setup. Each dataset is equally split into training and test sets.}
    \label{tab:multi_binary_compare}
    \tiny
    \setlength{\parindent}{-3em}
    \begin{tabular}{l | ccc | ccccc}
    \hline
    Prediction                                    & \multicolumn{3}{c|}{Multi-class (Acc (\%) $\uparrow$ $\lvert$ Macro-F1 $\uparrow$)}                & \multicolumn{5}{c}{Binary (Acc (\%) $\uparrow$ $\lvert$ ROC-AUC $\uparrow$ $\lvert$ AP $\uparrow$ )}  \\
    \hline
    Dataset                                       & \wiki(1k)                     & \wiki(5k)                        & \wiki(10k)                      & \wiki(10k)                                         & \twitter                                           & \ppi                                               & \dblp                                              & \blogcatalog                                           \\
    \hline                                                                                                                                                                                                                                                                                                                                                                                      
    AA~\cite{DBLP:journals/socnet/AdamicA03}      & 1.570$\lvert$0.031            &  0.330$\lvert$0.007              & 0.120$\lvert$0.003              &  49.98$\lvert$0.5493$\lvert$0.5498                 &  47.39$\lvert$0.6194$\lvert$0.5665                 &  50.30$\lvert$0.5605$\lvert$0.5583                 &  49.99$\lvert$0.6229$\lvert$0.6218                 &  49.51$\lvert$0.5776$\lvert$0.6099                     \\
    RA~\cite{zhou2009predicting}                  & 4.590$\lvert$0.088            &  0.210$\lvert$0.004              & 0.300$\lvert$0.001              &  49.98$\lvert$0.5499$\lvert$0.5499                 &  45.36$\lvert$0.6033$\lvert$0.5622                 &  50.17$\lvert$0.5605$\lvert$0.5583                 &  49.99$\lvert$0.6228$\lvert$0.6217                 &  49.54$\lvert$0.5747$\lvert$0.6064                     \\
    \hline                                                                                                                                                                                                                                                                                                                                                                      
    N2V~\cite{DBLP:conf/kdd/GroverL16}            & 51.53$\lvert$0.119            &  39.93$\lvert$0.065              & 35.46$\lvert$0.060              &  72.11$\lvert$0.7437$\lvert$0.8297                 &  64.98$\lvert$0.7059$\lvert$0.7052                 &  61.96$\lvert$0.7787$\lvert$0.7605                 &  88.04$\lvert$0.9587$\lvert$0.9535                 &  72.84$\lvert$0.8018$\lvert$0.8000                     \\
    S2V~\cite{DBLP:conf/kdd/RibeiroSF17}          & 43.83$\lvert$0.074            &  42.26$\lvert$0.035              & 42.96$\lvert$0.032              &  83.75$\lvert$0.9178$\lvert$0.9206                 &  69.16$\lvert$0.7473$\lvert$0.7140                 &  73.78$\lvert$0.8294$\lvert$0.8206                 &  67.23$\lvert$0.7381$\lvert$0.7464                 &  61.57$\lvert$0.6566$\lvert$0.6519                     \\
    HARP~\cite{DBLP:conf/aaai/ChenPHS18}          & 50.31$\lvert$0.120            &  53.05$\lvert$0.160              & 53.73$\lvert$0.156              &  89.81$\lvert$0.9447$\lvert$0.9189                 &  72.15$\lvert$0.7874$\lvert$0.7226                 &  64.68$\lvert$0.7218$\lvert$0.6718                 &  93.11$\lvert$0.9811$\lvert$0.9550                 &  77.07$\lvert$0.8568$\lvert$0.8494                     \\
    R2V~\cite{Luo2022ripple2vecNE}                & 26.28$\lvert$0.030	          &  33.15$\lvert$0.019              & 33.10$\lvert$0.023              &  62.16$\lvert$0.6548$\lvert$0.6221	                &  54.43$\lvert$0.5626$\lvert$0.5554                 &  66.32$\lvert$0.7240$\lvert$0.7543                 &  52.84$\lvert$0.5404$\lvert$0.5345                 &  60.27$\lvert$0.6355$\lvert$0.6036                     \\
    \hline                                                                                                                                                                                                                                                                                                                                                                                                          
    SEAL~\cite{DBLP:conf/nips/ZhangC18} & -- & -- & -- &  78.88$\lvert$0.8594$\lvert$0.8755                 &  76.81$\lvert$0.8325$\lvert$0.7750                 &  84.09$\lvert$0.8552$\lvert$0.8688                 &  81.37$\lvert$0.8538$\lvert$0.8927                 &  78.93$\lvert$0.8696$\lvert$0.8668                     \\
    RGCN~\cite{DBLP:conf/esws/SchlichtkrullKB18}  & 38.37$\lvert$0.088            &  48.99$\lvert$0.066              & 47.66$\lvert$0.082              &  50.00$\lvert$0.5000$\lvert$0.5000                 &  50.00$\lvert$0.5000$\lvert$0.5000                 &  50.80$\lvert$0.6978$\lvert$0.7056                 &  68.08$\lvert$0.7148$\lvert$0.7557                 &  69.10$\lvert$0.7605$\lvert$0.7476                     \\
    GCA~\cite{DBLP:conf/www/0001XYLWW21}          & 58.20$\lvert$0.228            &  56.71$\lvert$0.107              & 56.11$\lvert$0.094              &  89.03$\lvert$0.9539$\lvert$0.9601                 &  74.18$\lvert$0.8167$\lvert$0.7998                 &  73.03$\lvert$0.8374$\lvert$0.8298                 &  87.38$\lvert$0.9522$\lvert$0.9472                 &  69.79$\lvert$0.7659$\lvert$0.7582                     \\
    HGT~\cite{DBLP:conf/www/HuDWS20}              & 14.19$\lvert$0.013            &  12.04$\lvert$0.004              & 16.12$\lvert$0.003              &  50.60$\lvert$0.5063$\lvert$0.5058                 &  56.10$\lvert$0.5928$\lvert$0.5818                 &  53.96$\lvert$0.5501$\lvert$0.5358                 &  50.47$\lvert$0.5029$\lvert$0.5091                 &  50.06$\lvert$0.5036$\lvert$0.5030                     \\
    GAug~\cite{DBLP:conf/aaai/0003LNW0S21}        & 18.33$\lvert$0.006            &  21.96$\lvert$0.004              & 18.38$\lvert$0.002              &  50.00$\lvert$0.5000$\lvert$0.5000                 &  50.00$\lvert$0.5000$\lvert$0.5000                 &  50.85$\lvert$0.6998$\lvert$0.7085                 &  68.42$\lvert$0.7199$\lvert$0.7561                 &  69.32$\lvert$0.7620$\lvert$0.7489                     \\
    BUDDY~\cite{chamberlain2022graph}             & 60.59$\lvert$0.571            &  61.21$\lvert$\ct{0.589}         & 63.41$\lvert$0.611              &  90.67$\lvert$0.9776$\lvert$0.9803                 &  82.58$\lvert$0.9127$\lvert$0.8997                 &  70.65$\lvert$0.7773$\lvert$0.8066                 &  89.74$\lvert$0.9748$\lvert$0.9767                 &  82.68$\lvert$0.9121$\lvert$0.9140                     \\
    NCN~\cite{wang2024neural}                     & 46.34$\lvert$0.075            &  47.93$\lvert$0.046              & 55.97$\lvert$0.067              &  90.36$\lvert$0.9773$\lvert$0.9762                 &  83.29$\lvert$\ct{0.9219}$\lvert$\ct{0.9097}       &  61.69$\lvert$0.7901$\lvert$0.8108                 &  86.89$\lvert$0.9665$\lvert$0.9700                 &  77.07$\lvert$0.8892$\lvert$0.8879                     \\
    NCN-diff~\cite{wang2024neural}                & 45.71$\lvert$0.071            &  50.78$\lvert$0.082              & 61.43$\lvert$0.133              &  90.23$\lvert$0.9772$\lvert$0.9735                 &  59.88$\lvert$0.9153$\lvert$0.8996                 &  59.78$\lvert$0.8050$\lvert$0.8187                 &  88.20$\lvert$0.9694$\lvert$0.9714                 &  80.32$\lvert$0.8944$\lvert$0.8926                     \\
    NCNC~\cite{wang2024neural} & -- & -- & -- &  94.45$\lvert$0.9851$\lvert$0.9856                 &  83.54$\lvert$0.9182$\lvert$0.9084                 &  68.17$\lvert$0.8216$\lvert$0.8471                 &  86.74$\lvert$0.9673$\lvert$0.9713                 &  80.06$\lvert$0.8874$\lvert$0.8865                     \\
    Refined-GAE~\cite{ma2025reconsidering}        & 48.94$\lvert$0.090            &  45.09$\lvert$0.239              & 38.53$\lvert$0.115              &  89.90$\lvert$0.9591$\lvert$0.9568                 &  72.47$\lvert$0.8688$\lvert$0.8228                 &  60.88$\lvert$0.8039$\lvert$0.8061                 &  94.09$\lvert$0.9807$\lvert$0.9840                 &  84.42$\lvert$0.9237$\lvert$0.9186                     \\
    \hline                                                                                                                                                                                                                                                                                                                                                                                                                                              
    \framework (Ours)                             & \ct{78.75}$\lvert$\ct{0.576}  &  \ct{81.58}$\lvert$0.535         & \ct{87.38}$\lvert$\ct{0.625}    &  \ct{95.22}$\lvert$\ct{0.9879}$\lvert$\ct{0.9859}  &  \ct{84.51}$\lvert$0.9140$\lvert$0.8930            &  \ct{90.88}$\lvert$\ct{0.9641}$\lvert$\ct{0.9571}  &  \ct{95.33}$\lvert$\ct{0.9934}$\lvert$\ct{0.9938}  &  \ct{92.27}$\lvert$\ct{0.9772}$\lvert$\ct{0.9713}      \\
    Relative Gain (\%)                            & +30.0$\lvert$+0.88            &  +33.3$\lvert$-9.17              & +37.8$\lvert$+2.29              &  +0.82$\lvert$+0.28$\lvert$+0.03                   &  +1.16$\lvert$-0.86$\lvert$-1.84                   &  +8.07$\lvert$+12.7$\lvert$+10.2                   &  +1.32$\lvert$+1.25$\lvert$+1.00                   &  +9.30$\lvert$+5.79$\lvert$+5.74                       \\
    \hline
    \end{tabular}
}

\normalfig[t]{
\includegraphics[scale=0.5]{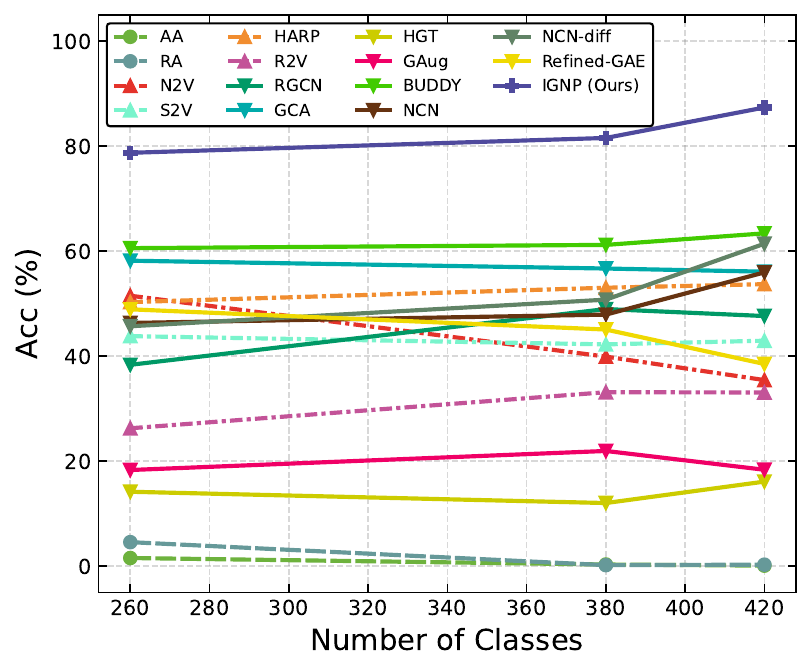}
\caption{Accuracy for multi-class prediction on \textit{Wikidata} with 1k, 5k, and 10k entities.
The number of classes for prediction expands in proportion to the entity size. All baselines are re-evaluated and fine-tuned under our unified experimental setup.}
\label{fig:multi_compare_accuracy}
}

\subsection{Multi-class prediction}\label{subsec:multi-class-prediction}
The accuracy and macro-F1 for 1k, 5k, and 10k nodes from \wiki with about 260, 380, and 420 edge types for multi-class link prediction are shown in \tabref{tab:multi_binary_compare} (edge types are not considered by SEAL and NCNC), and the trend with the number of classes increases is in \figref{fig:multi_compare_accuracy}.
We observe that \framework outperforms all baselines by 30\%--37\% and consistently achieves the best accuracy, which further improves as the number of classes increases.
For \framework, relevant local and global information from both $\hout$ and $\hin$ is well utilized by the sub-graph with virtual edges.
Additionally, more data provides diversified sub-graphs, which bring additional supervision signals of each class to improve the results even as the number of classes increases.

In general, most latent feature methods and GNN-based approaches outperform traditional similarity metrics. 
Specifically, methods such as AA, RA, N2V, S2V, RGCN, GAug, GCA, and Refined-GAE primarily rely on local node features and topological information from $\hout$. 
As the number of relations increases, the prediction of these methods becomes more challenging, resulting in sub-optimal performance. 
Notably, the performance of HARP, R2V, BUDDY, NCN, and NCN-diff improves with larger datasets due to their ability to implicitly or explicitly incorporate global information from higher orders of neighbors to refine local embeddings. 
However, this process is less efficient and may introduce additional noise.

\subsection{Binary prediction}\label{subsec:binary-prediction}
The accuracy, ROC-AUC, and AP results for binary prediction of edge existence are shown in \tabref{tab:multi_binary_compare}.
We observe that \framework achieves the best accuracy on all datasets, outperforming the baselines by up to 9\%.
Our \framework enriches the local features with real long-distance connectivity via global graph searching to improve performance, which especially benefits the results on \wiki (10k) where nodes across the entire graph are frequently connected.
Edges in \twitter, \dblp, \blogcatalog, and \ppi record interactions among a group of people or proteins from parts of the graph, which can be better predicted with global information.
Besides, results from \twitter indicate that \framework can also learn to effectively predict the existence of edges using observable attributes of the nodes.

Meanwhile, focusing on local topology, the performance of similarity metrics, RGCN, and SEAL is limited.
HARP utilizes global information from $\hout \cup \noticedset()$ with $\Q \to \infty$ and outperforms other latent feature methods, including N2V, R2V, and S2V, demonstrating its capability to reconstruct edge existence in graphs with stable patterns, such as \ppi, where proteins within the same category exhibit numerous shared properties.
GNN-based baselines generally outperform latent feature methods on different datasets.
Compared with baselines of latent feature methods and similarity metrics, NCN, NCN-diff, and NCNC are better on \wiki (10k) and \twitter; SEAL is better on \ppi, and Refined-GAE is better on larger datasets including \dblp and \blogcatalog.
GCA and GAug augment the data with random noise to obtain a more general model, during which the model might also be confused by the induced noise.

\widetab[tb]{
\caption{Accuracy comparison of different sub-graph structures, loss functions, and order of neighbors. Each dataset is equally split into training and test sets.}
\label{tab:accuracy_bi}
\scriptsize
\begin{tabular}{c|c|c|ccc|ccc|ccc}
    \hline
Acc (\%) $\uparrow$    & -        &&& \textit{CB} &&& \textit{PL} &&& \framework & ($\Q=5$) \\
    \hline
    Task        & Dataset    & $\nborder$($\nborder[\ast]$)& 0 & 1 & 2 & 0 & 1 & 2 & 0 (1) & 1 (6) & 2 (12)      \\
    \hline
                &            & $\loss(0)$ & 88.24        & \emph{93.20} & 86.12 & 88.20        & 94.77        & 93.09 & 90.55          & 94.82          & 93.74   \\
    Binary      & \wiki      & $\loss(1)$ & 89.46        & 92.52        & 89.02 & 90.37        & \emph{94.98} & 93.70 & 90.80          & \textbf{95.22} & 94.22   \\
                &  (10k)     & $\loss(2)$ & 62.14        & 71.42        & 68.90 & 64.78        & 69.19        & 68.27 & 69.45          & 64.49          & 70.22   \\
    \hline                                                                                                                                                 
                &            & $\loss(0)$ & \emph{81.00} & 75.77        & 79.42 & \emph{81.16} & 71.37        & 50.84 & \textbf{81.99} & 64.22          & 74.24  \\
    Binary      & \twitter   & $\loss(1)$ & 80.39        & 76.62        & 78.93 & 80.10        & 70.94        & 49.60 & 80.36          & 61.62          & 73.81  \\
                &            & $\loss(2)$ & 58.08        & 59.97        & 51.40 & 55.64        & 63.42        & 68.70 & 61.41          & 56.57          & 72.21  \\
    \hline                                                                                                                                                                 
                &            & $\loss(0)$ & \emph{73.77} & 65.15        & 64.42 & 67.27        & 75.61        & 70.60 & 71.63          & 75.88          & 68.77  \\
    Binary      & \ppi       & $\loss(1)$ & 71.99        & 66.07        & 62.85 & 73.45        & \emph{79.54} & 73.30 & 73.76          & \textbf{90.88} & 68.99  \\
                &            & $\loss(2)$ & 58.95        & 61.30        & 63.52 & 59.00        & 65.06        & 56.50 & 58.62          & 58.80          & 58.36  \\
    \hline                                                                                                                                                                        
                &            & $\loss(0)$ & 76.26        & 70.36        & 50.21 & 78.34        & 82.01        & 79.89 & 77.08          & 79.24          & 74.15  \\
    Multi-class & \wiki      & $\loss(1)$ & \emph{78.84} & 68.19        & 51.83 & 78.76        & \emph{83.20} & 78.71 & 72.51          & \textbf{81.63} & 79.69  \\
                &  (10k)     & $\loss(2)$ & 29.31        & 32.93        & 32.98 & 33.12        & 31.06        & 39.54 & 36.08          & 33.56          & 35.97  \\
    \hline
\end{tabular}
}

\subsection{Method analysis and ablation study}\label{subsec:ablation-study}

\subsubsection{Effectiveness of inward influence}\label{subsubsec:inward-influence}
To show the performance when utilizing inward influence from nodes defined in \theref{the:H_in}, \ie relevant global information, we compare the accuracy of \framework and its variants, which use different sub-graph structures, shown in \tabref{tab:accuracy_bi}.
Let $\nborder[\ast]$ be the max order of utilized neighbors, and then $\nborder[\ast]$ of \framework, \textit{CB}, and \textit{PL} will be~$\max \rbkt{n \cdot \rbkt{\Q + 1}, 1}$, $n$, and $n$ respectively, where $\nborder$ is the order of neighbors of the sub-graphs.
We can find that \framework achieves the best results in all cases for binary and multi-class prediction.

Since \textit{CB} ignores the neighboring graph structure, it prefers very small $\nborder$ (\ie $\nborder = 0$) to reduce information, and the performance is therefore capped.
We can see \textit{CB} reaches its best on \twitter and \ppi when $\nborder = 0$, and it cannot handle more information when $\nborder > 0$.

By considering the graph structure of nodes in $\hout$, the performance of \textit{PL} increases dramatically compared with the results of \textit{CB}.
Although compared with \framework, \textit{PL} shares the same GNN network structure and hyperparameters, the absence of information in virtual edges capped the performance of \textit{PL}.

To summarize, using the local structure information from $\hout$ improves the performance, as shown by \textit{CB} and \textit{PL}.
Nonetheless, we find that only using $\hout$ is not sufficient, and a larger range of neighborhoods with relevant global information should also be considered, such as nodes in $\hin$.
The virtual edges provide more global information in $\hin$.
By combining $\hout$, more supervision signals will improve the \framework for link prediction.

\subsubsection{Effectiveness of virtual edges}
To compare the effectiveness of virtual edges, we compared the accuracy of binary prediction using the entire sub-graph of influence propagation (\framework-NoVE), virtual sub-graph (\framework), and plain sub-graph (\textit{PL}). Results of different $\Q$ after 20 epochs of training are shown in \tabref{tab:effectiveness_ve}.
We set $\nborder[\ast]$ as 2, 3, 6, 11, thus \textit{PL} uses $\nborder = 2, 3, 6, 11$ to generate sub-graphs whereas \framework constantly uses $\nborder=1$ with $\Q = 1, 2, 5, 10$.
\normaltab[tb]{
\caption{The binary prediction accuracy using \framework, \framework-NoVE, and \textit{PL} under different $\nborder[\ast]$ on \wiki (10k) with equally split training/test sets.}
\label{tab:effectiveness_ve}
\begin{tabular}{l rrrr}
    \toprule
    Acc (\%) $\uparrow$                     & $\nborder[\ast] = 2$ & $\nborder[\ast] = 3$ & $\nborder[\ast] = 6$ & $\nborder[\ast] = 11$ \\
                                 & $\rbkt{\Q = 1}$         & $\rbkt{\Q = 2}$      & $\rbkt{\Q = 5}$     &  $\rbkt{\Q = 10}$ \\
    \midrule
    \framework                   & 94.93           & 95.45        & 95.09       & 95.05         \\
    \framework-NoVE              & 95.33           & 95.41        & 94.84       & 95.20         \\
    \midrule
    \textit{PL}                  & 93.50           & 92.88        & 91.35       & 87.99         \\
    \bottomrule
\end{tabular}
}
We can see from the results that the accuracy of using and not using the virtual edges is almost identical, but with a slight performance regression.
It is a fact that both of them utilize the information from the sub-graph of influence propagation, and the use of virtual edges does not lose a lot of global information for link prediction.
Thus, the virtual edge does not considerably affect the performance.
From the result of \textit{PL} when $\nborder[\ast] = 2$ and the result of \framework when $\nborder[\ast] = 3$ to $\nborder[\ast] = 11$, we can see a 1.5 to 2.0 percentage performance improvement, which indicates the inward influence is beneficial for the link prediction.
Besides, due to the irrelevant noise, the performance of \textit{PL} becomes worse compared with \framework and \framework-NoVE.

\subsubsection{Efficiency of virtual edge}
To show the efficiency advantage of using virtual edges, we fix the $\nborder$ of \framework and change $\Q$ to utilize a larger range of nodes.
We set $\nborder[\ast]$ as 2, 3, 6, 11, thus \textit{PL} uses $\nborder = 2, 3, 6, 11$ to generate sub-graphs whereas \framework constantly uses $\nborder=1$ with $\Q = 1, 2, 5, 10$.
For fairness, \textit{PL} and \framework use the same settings to train the model for 50 epochs without the early stop.
The execution (wall clock) time of sub-graph generation, GNN model training, and overall procedure of \framework and \textit{PL} for binary prediction are shown in \figref{fig:efficiency_ve}.
\normalfig[t]{
\includegraphics[scale=0.55]{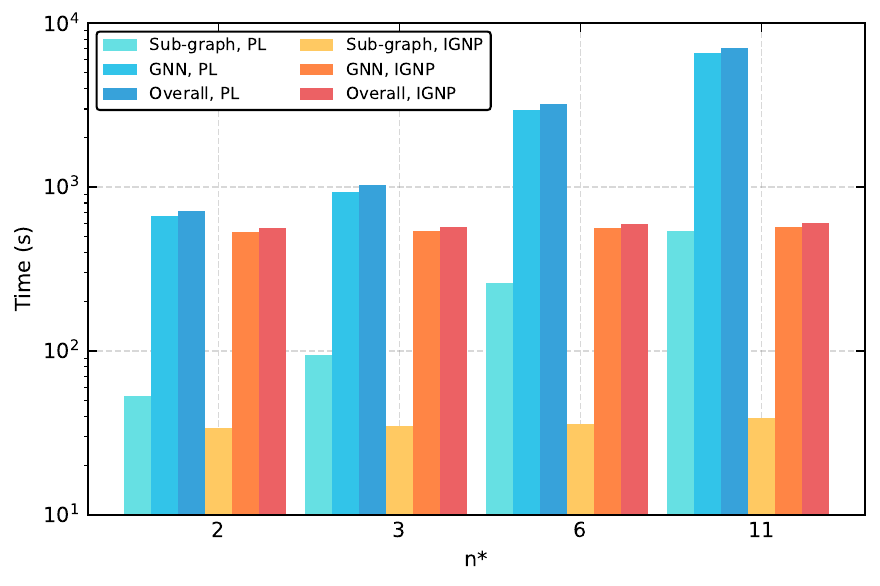}
\caption{The execution time of \framework and \textit{PL} for binary prediction under different $\nborder[\ast]$ on \wiki (10k).}
\label{fig:efficiency_ve}
}

The execution time of \textit{PL} is increasing exponentially with $\nborder[\ast]$ becoming larger, which gives the practical evidence for \lemref{the:complexity_of_GNNs} in \secref{subsec:analysis}.
On the contrary, the execution time of \framework increases slightly and approximately satisfies \theref{the:ve_speedup} compared with \textit{PL}.
Although searching a larger range, the guided traverse in \agoref{alg:extract_the_virtual_sub-graph} reduces the time of generating \emph{virtual} sub-graphs.
Due to the sub-graph size not considerably changing, the training time of \framework slowly increases.
Overall, \framework shows considerable time efficiency compared with common GNN-based link prediction methods when learning the feature.

\normaltab[t]{
    \caption{Accuracy of different size of entities for binary prediction of \framework using $\loss(1)$. Each dataset is equally split into training and test sets.}
    \label{tab:accuracy_trend_bi}
    \begin{tabular}{c|ccc}
    \toprule
    Acc (\%) $\uparrow$ & $\nborder=0$ & $\nborder=1$ & $\nborder=2$ \\
    \midrule
    \wiki (1k)   & 80.61 & \textbf{87.75} & 82.35 \\
    \wiki (5k)   & 86.62 & \textbf{92.75} & 90.40 \\
    \wiki (10k)  & 90.80 & \textbf{95.22} & 94.22 \\
    \bottomrule
    \end{tabular}
}

\normalfig[tb]{
\includegraphics[scale=0.55]{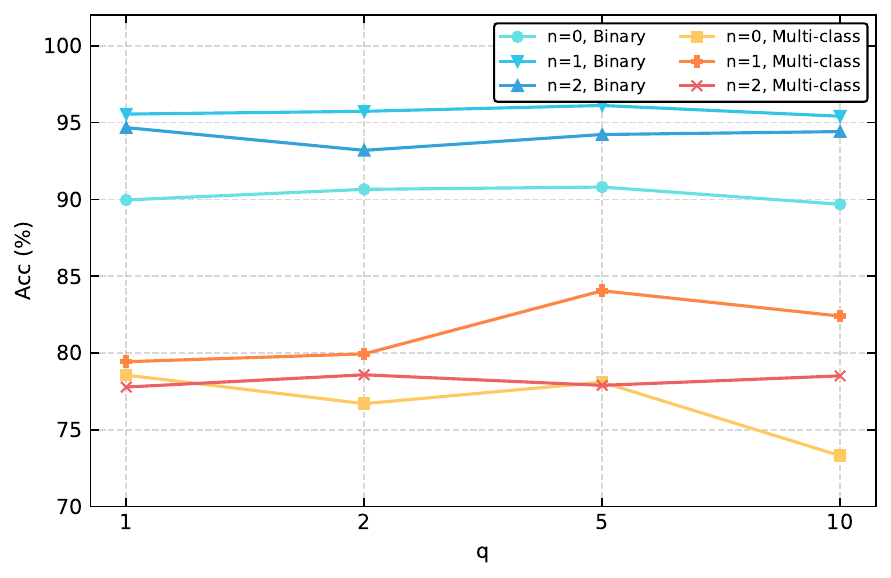}
\caption{Accuracy comparison of different $\Q$ and $\nborder$ on \wiki (10k). }
\label{fig:accuracy_wikidata_10k_Q}
}

\subsubsection{Comparison of different losses}
\tabref{tab:accuracy_bi} shows how different loss functions defined in \secref{subsec:link-prediction-using-a-GNN} affect the performance of different sub-graph structures and datasets.
We can find $\loss(0)$ is better on \twitter, $\loss(1)$ is better in most cases on \wiki (10k) and \ppi, and $\loss(2)$ is the worst loss.

That $\loss(0)$ is better on \twitter is probably because neighbor information is not very important, and the model should concentrate on the selected pair of nodes.
On \twitter, the ego user has the edge over all of his friends, but few friends of him have the edge over each other.
Using $\loss(0)$ with $\nborder = 0$ to ignore the neighborhood information will make the performance generally better on \twitter, which is in accordance with the results.
For \textit{CB}, $\loss(0)$ is better on \ppi because \textit{CB} cannot utilize neighborhood information well.

$\loss(1)$ focuses on the interaction among the sub-graph.
It treats all the influences within a sub-graph comprehensively, and different patterns of influence propagation can be recognized.
\wiki (10k) and \ppi have more interaction information, thus $\loss(1)$ works well with most cases and obtains better results.

$\loss(2)$ only utilizes labels of $\edgeij$, which is not enough to learn all the patterns of influence propagation.

\subsubsection{Parameter analysis}
\parahead{Varying $\nborder$ for binary and multi-class prediction.}
A larger $\nborder$ does not guarantee a better result according to \tabref{tab:accuracy_bi}.
Using \textit{CB} with $\nborder = 0$ and using \textit{PL}, \framework with $\nborder = 1$ is more likely to produce good results.
Note that when $\nborder = 0$, the difference between the structure of \textit{CB} and \textit{PL} will be the loop edge of $\nodei$, and that between \textit{PL} and \framework will be the virtual edges.
When $\nborder > 0$, neighborhood information can be incorporated to boost the performance of \framework and \textit{PL}. Nevertheless, an overly large $\nborder$ introduces excessive irrelevant information and thus degrades performance.
This trend is evaluated with the first 1k, 5k, and 10k entities of \wiki under different $\nborder$, in \tabref{tab:accuracy_trend_bi}.

\normalfig[tb]{
\subfloat[Binary Loss ($\nborder=0$)]{
    \includegraphics[scale=0.35]{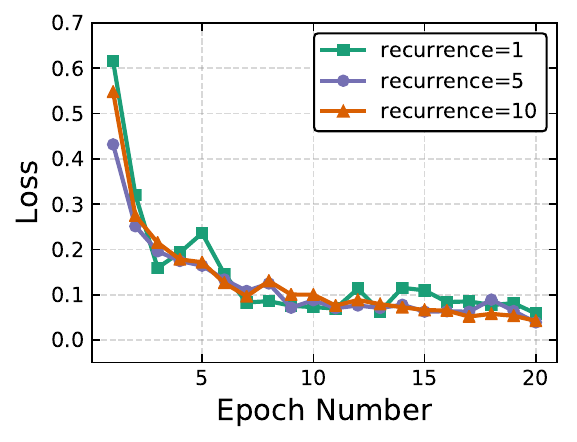}
    \label{fig:ltr_order_0_bi}
} 
\subfloat[Binary Acc ($\nborder=0$)]{
    \includegraphics[scale=0.35]{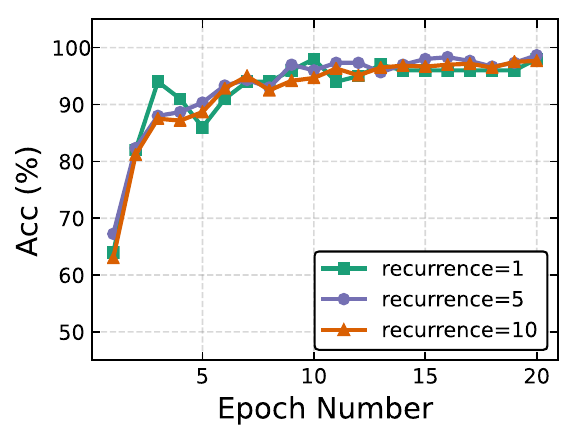}
    \label{fig:accuracy_order_0_bi}
} 
\subfloat[Multi-class Loss ($\nborder=0$)]{
    \includegraphics[scale=0.35]{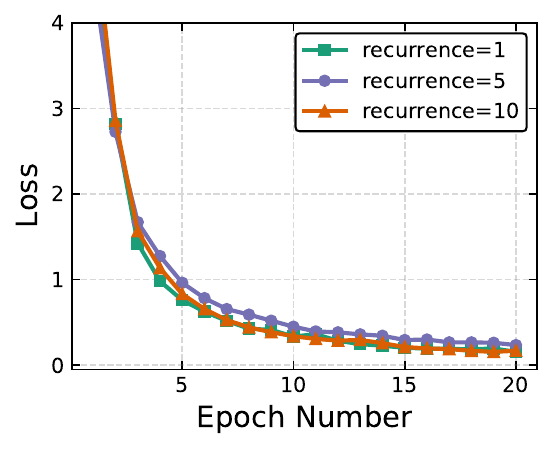}
    \label{fig:ltr_order_0_multi}
}
\subfloat[Multi-class Acc ($\nborder=0$)]{
    \includegraphics[scale=0.35]{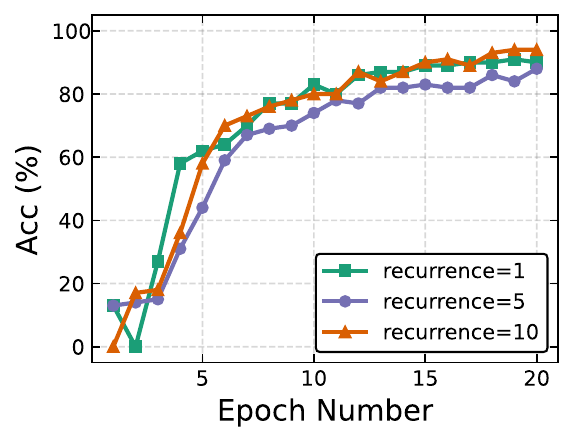}
    \label{fig:accuracy_order_0_multi}
} \\
\subfloat[Binary Loss ($\nborder=1$)]{
    \includegraphics[scale=0.35]{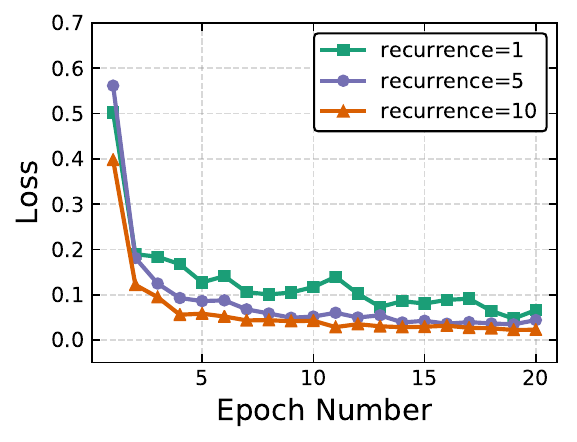}
    \label{fig:ltr_order_1_bi}
}
\subfloat[Binary Acc ($\nborder=1$)]{
    \includegraphics[scale=0.35]{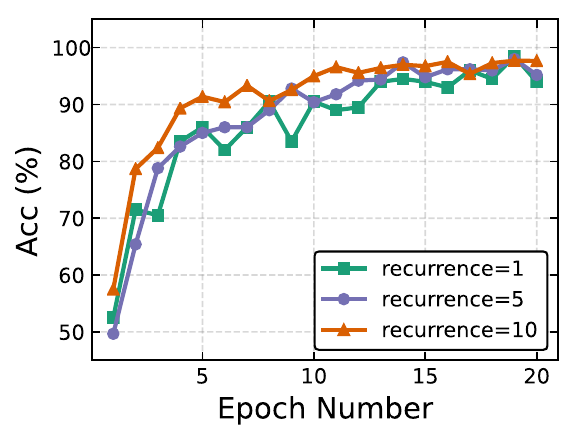}
    \label{fig:accuracy_order_1_bi}
} 
\subfloat[Multi-class Loss ($\nborder=1$)]{
    \includegraphics[scale=0.35]{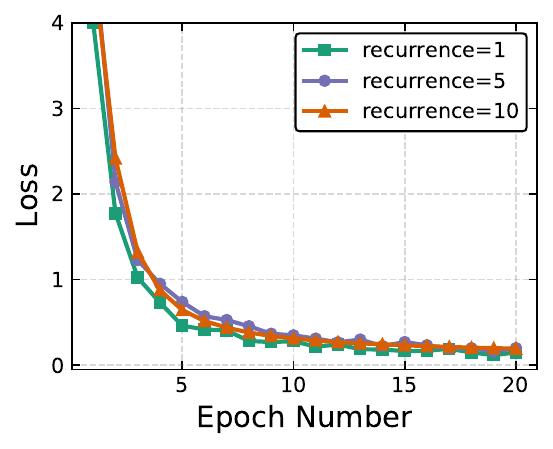}
    \label{fig:ltr_order_1_multi}
}
\subfloat[Multi-class Acc ($\nborder=1$)]{
    \includegraphics[scale=0.35]{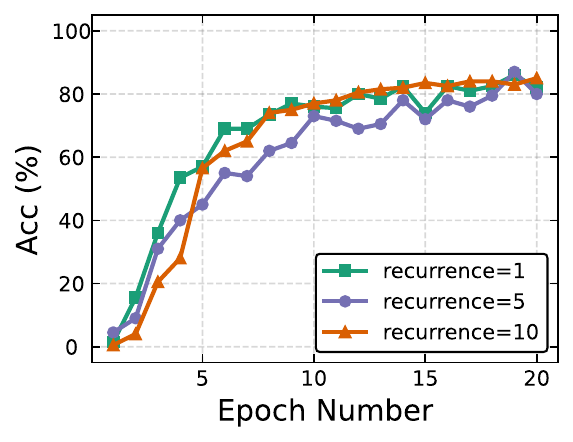}
    \label{fig:accuracy_order_1_multi}
} \\
\subfloat[Binary Loss ($\nborder=2$)]{
    \includegraphics[scale=0.35]{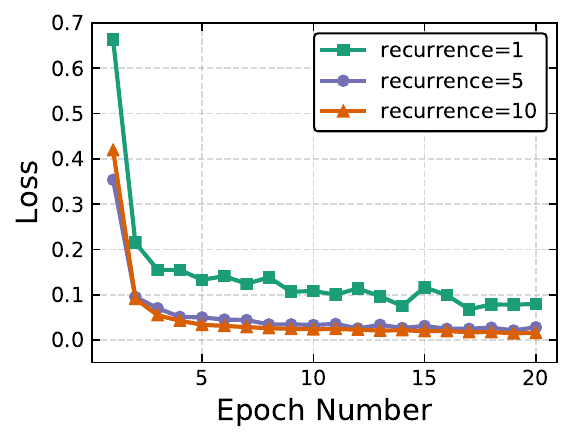}
    \label{fig:ltr_order_2_bi}
}
\subfloat[Binary Acc ($\nborder=2$)]{
    \includegraphics[scale=0.35]{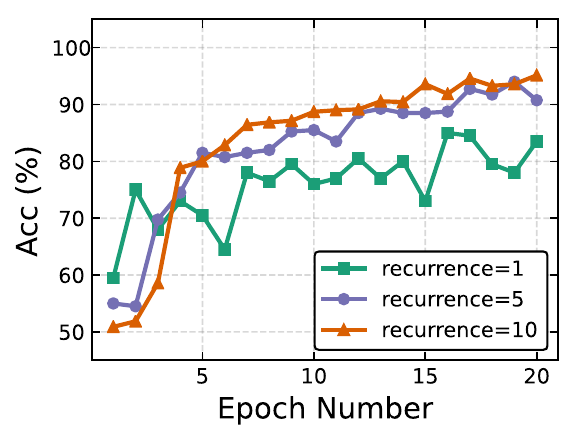}
    \label{fig:accuracy_order_2_bi}
} 
\subfloat[Multi-class Loss ($\nborder=2$)]{
    \includegraphics[scale=0.35]{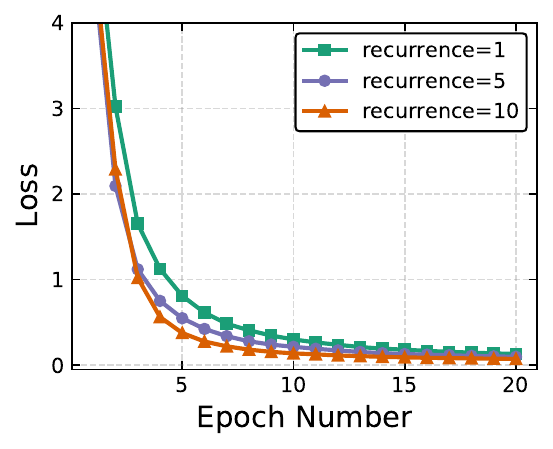}
    \label{fig:ltr_order_2_multi}
}
\subfloat[Multi-class Acc ($\nborder=2$)]{
    \includegraphics[scale=0.35]{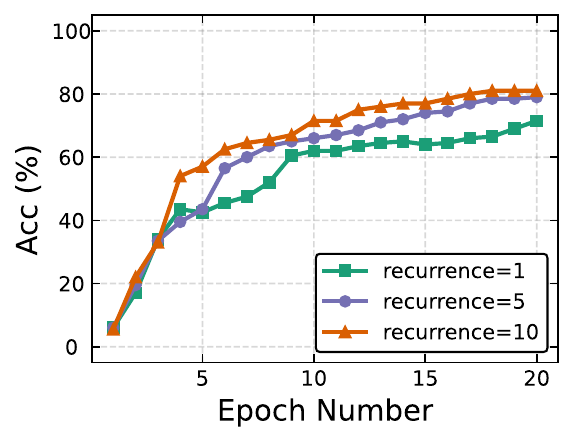}
    \label{fig:accuracy_order_2_multi}
} 
\caption{Training loss and accuracy curves on \wiki (10k). The model achieves a slighter lower loss (higher accuracy) with $\recurrence = 10$.}
\label{fig:r_on_0_2}
}

\parahead{Varying $\Q$ on \wiki (10k).}
We show the comparison of different $\Q$ under different $\nborder$ in \figref{fig:accuracy_wikidata_10k_Q}.
We observe a clear optimum for the values $\Q=5$ and $\nborder=1$ in both binary and multi-class prediction.

\parahead{Message passing iterations.}
We examine the training accuracy of \framework ($\Q=5$) with $\loss(1)$ loss on \wiki (10k) with the different number of iterations, \ie recurrence $\recurrence$, of the GNN in \figref{fig:r_on_0_2}.
As we can observe in binary prediction, the model with a larger $\recurrence$ converges faster than those with a smaller $\recurrence$ and achieves a slightly better accuracy after 20 epochs of training. Such a trend continues after that.
The same observation holds for varying values of $\nborder$.
As to multi-class prediction, a smaller $\recurrence$ performs better at first, but the advantage decreases and disappears after about 15 epochs of training, and the accuracy for $\recurrence = 10$ surpasses other settings after 20 epochs of training.

\subsubsection{Visualization of training}
\figref{fig:visualization} shows the advantages of using the virtual sub-graph to generate edge embeddings by utilizing more information from the global structure.
Predicting the types of edges will be easier given more global information, which helps the model generate more distinguishable edge embeddings as shown in \figref{fig:sub-graphs}.
Given the virtual sub-graphs, \framework learns to predict the types of all the edges, and the performance rapidly increases during the training process, shown in \figref{fig:training-process}.
\normalfig[tb]{
\subfloat[Different sub-graphs and their edge embeddings]{
\includegraphics[scale=0.275]{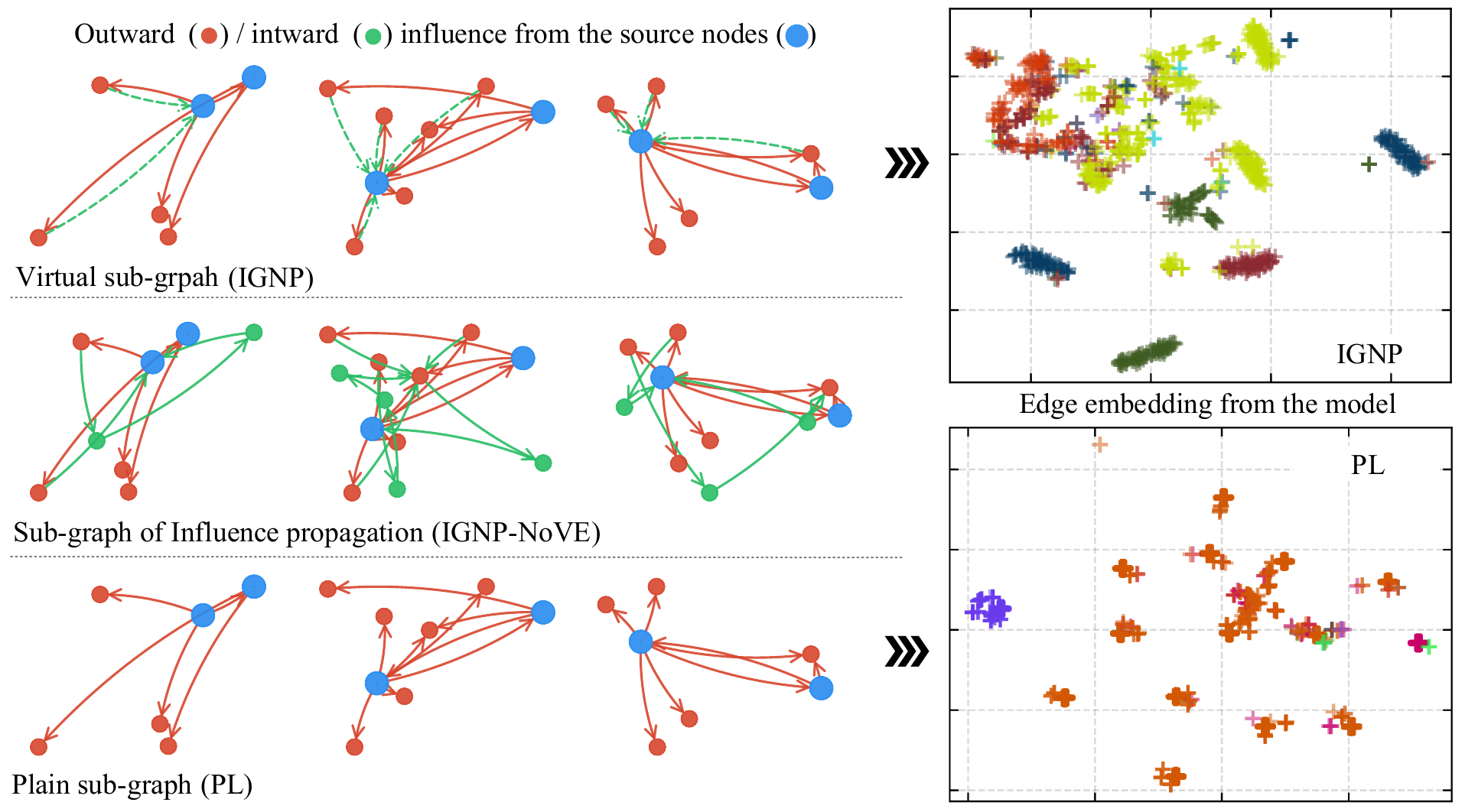}
\label{fig:sub-graphs}
}\\
\subfloat[Training results after 2, 10, and 20 epochs and original graph]{
\includegraphics[scale=0.5]{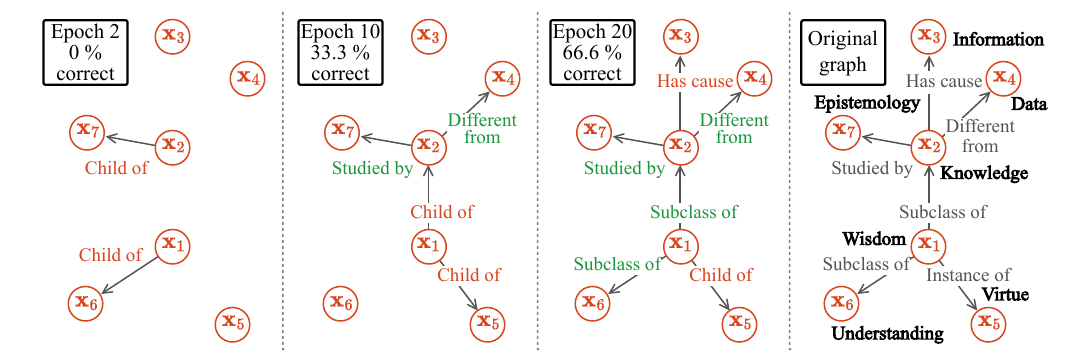}
\label{fig:training-process}
}
\caption{Visualization of the training for multi-class prediction on \wiki (10k).}
\label{fig:visualization}
}

\section{Conclusions}\label{sec:conclusions}

In this study, we model relevant information for link prediction through the lens of node influence, and enclose the influence between any pair of nodes within sub-graphs by extending the classical SIR propagation model. 
We further compress the influences into much smaller sub-graphs by the virtual edges, derived from mean-field theory, to significantly reduce the computation when using the global graph structure.
These compressed structures are then utilized to extract relevant local and global information for link prediction. 
Our analysis reveals a strong correlation between node influence and the relevant information for link prediction, as substantiated by extensive empirical evaluations. 
The proposed \framework framework consistently outperforms representative baselines across multiple benchmark datasets, offering novel insights into link prediction through the paradigm of influence propagation.

A limitation of our \framework is that it is designed to predict relations within static graphs. 
Nonetheless, the dynamics of relational changes, a common occurrence in numerous real-world applications, remain unaddressed by the \framework framework.
The node influence propagation is also applicable in dynamic scenarios, as influence continuously spreads over time.
Consequently, building upon the findings of our investigation, the prediction of relations within dynamic graphs represents a potential direction for future research.

\section*{CRediT authorship contribution statement}
\textbf{Zidu Yin}: Writing -- original draft, Visualization, Validation, Software, Methodology, Investigation, Formal analysis, Conceptualization.
\textbf{Yuankai Qi}: Writing -- original draft, Methodology, Investigation.
\textbf{Dong Gong}: Writing -- original draft, Methodology, Investigation.
\textbf{Ehsan Abbasnejad}: Writing -- original draft, Investigation.
\textbf{Kun Yue}: Writing -- original draft, Supervision, Funding acquisition.
\textbf{Javen Qinfeng Shi}: Writing -- original draft, Supervision, Project administration, Methodology, Investigation.

\section*{Declaration of competing interest}
The authors declare that they have no known competing financial interests or personal relationships that could have appeared to influence the work reported in this paper.

\section*{Acknowledgments}
This work was supported by the Joint Key Project of the National Natural Science Foundation of China (U23A20298), the Yunnan Fundamental Research Project (202301AU070147), and the Open Project Program of Yunnan Key Laboratory of Intelligent Systems and Computing (ISC24Y08).
Yuankai Qi is not supported by the aforementioned funds.

\section*{Data availability}
Data will be made available on request.


\appendix

\section{Details of baselines}\label{sec:baselines}
The official code or widely recognized implementations, based on the original papers, are adopted for all baselines. 
To be specific, we use the official code for baselines including N2V, S2V, LINE, HARP, SEAL, GAug, HGT, RGCN, GCA, BUDDY, NCN, NCN-diff, NCNC, and Refined-GAE. 
AA and RA are implemented by NetworkX,\footnote{\url{https://networkx.org/}} the software for complex networks.
All baselines were first reproduced the original results on their respective datasets reported in their original papers if applicable, before adapting them to our experimental setup.
Each datasets are pre-processed identically for different baselines, and the split of the datasets to train and evaluate the model is the same.
Meanwhile, the grid search was conducted on the most critical hyperparameters of different baselines to ensure we obtained the best results for all baselines. 
More detailed hyperparameters of different baselines are listed in \tabref{tab:hyperparams}.
\begin{table}[h]
\centering
\small
\caption{Key hyperparameters of baseline methods.}
\label{tab:hyperparams}
\begin{tabular}{l p{0.68\linewidth}}
\toprule
\textbf{Method} & \textbf{Key hyperparameters} \\
\midrule
N2V, S2V     & Default hyperparameters (yield the best performance). \\
HARP         & Node feature model: N2V. \\
SEAL         & Neighbor auto-detection enabled. \\
GAug         & $\alpha$=0.13; $\beta$=3.2; temperature=1.0. \\
HGT          & Layers=2; heads=4. \\
RGCN         & Dropout=0.2; lr=1e\textminus 2; grad norm=1.0; bases=4. \\
GCA          & lr=1e\textminus 2; hidden=256; activation=prelu; drop edge rates 1/2=0.3/0.4; $\tau$=0.4. \\
BUDDY        & lr=1e\textminus 4 (no decay); hidden channels=1024; epochs=100. \\
NCN/NCN-diff/NCNC & GNN dropout=0.7; predictor dropout=0.05; $\alpha$=1.0. \\
Refined-GAE  & lr=1e\textminus 3; GNN layers=4; dropout=0.6. \\
\bottomrule
\end{tabular}
\end{table}

\section{Proof of \texorpdfstring{\theref{the:H_out}}{Theorem 1}}\label{sec:proof-of-theorem-h-out}
\pf{
    Let $\propEvt(i)$ be the event for $\nodei$ been influenced by $\source$, and $\propEvt(i) \rbkt{k}$ be the times for $\nodei$ been influenced by $\source$ after $k$ independent propagations.
    Given the fact that the edges are from the influences of the node, but not all the edges are from the influence of $\source$.
    The influence of $\source$ will propagate over the existing edges with a probability $\infprob{ij}$, and the expectation of $\propEvt(i)$ will be
    \begin{equation}
        \label{eq:E_propEvt}
        \displaystyle \E \bbkt{\propEvt(i)} = \sum_{\singlepath \in \paths(si)} \prod_{\edge(ab) \in \singlepath} \infprob{ab} \period
    \end{equation}
    Note that the total number of the influenced nodes will be the $\influenced$ nodes of $\source$ at $\tsp$, \ie $\mR[\tsp](s)\rbkt{\nodeset}$.
    $\mR[\tsp](s)\rbkt{\nodeset} \cdot \E \bbkt{\propEvt(i)}$ indicates the number of influenced nodes $\nodei$ contributes.
    If $\mR[\tsp](s)\rbkt{\nodeset} \cdot \E \bbkt{\propEvt(i)} \ge 1$, $\nodei$ will be influenced by $\source$.
    Thus, let $\widehat{\hout}$ be the expected set of influence of $\source$ which is $\eqcbkt{ \nodei: \sum_{\singlepath \in \paths(si)} \prod_{\edge(ab) \in \singlepath} \infprob{ab} \ge \frac{1}{\mR[\tsp](s)\rbkt{\nodeset}} }$.
    On the other hand, $\lim_{k \rightarrow \infty} \E \eqbbkt{ \eqabs{\hout \setminus \widehat{\hout}} } = 0$.
    Thus, $\hout$ will be converged in probability to $\widehat{\hout}$.
}

\section{Proof of \texorpdfstring{\cororef{the:H_out_approx}}{Corollary 1}}\label{sec:proof-of-h-out-approx}
To prove \cororef{the:H_out_approx}, we first introduce the Radon--Nikodym Theorem.
\lemm[Radon--Nikodym Theorem~\cite{stein2009real}]{
    \label{the:radon_nikodym_theorem}
    Given a measurable space $\rbkt{ \nodeset, \sigfd, \mInf(s) \rbkt{\cdot} }$ and a set $\randsubset \in \sigfd$, there will be a function $\mr[\ts](s) \rbkt{\cdot}$ of the $\influenced$ nodes on $\randsubset$ such that
    \begin{equation}
        \displaystyle \mR[\ts](s) \rbkt{\randsubset} = \sum_{i \in \randsubset} \mr[\ts](s) \rbkt{i} \mInf[\ts](s) \rbkt{i} \period
        \label{eq:R_r_relation}
    \end{equation}
}
Integrating over the entire node set $\nodeset$ using \lemref{the:radon_nikodym_theorem}, we have
\begin{equation}
    \displaystyle \mR[\ts](s) \rbkt{\nodeset} = \sum_{i \in \nodeset} \mr[\ts](s) \rbkt{i} \mInf[\ts](s) \rbkt{i} \period
\end{equation}
Note that $\mR[\ts](s) \rbkt{\nodeset} = \sum_{i \in \nodeset} \frac{\mR[\ts](s) \rbkt{i}}{\mInf[\ts](s) \rbkt{i}} \mInf[\ts](s) \rbkt{i}$, and  thus
\begin{equation}
    \displaystyle \mr[\ts](s) \rbkt{\nodeset} = \frac{\mR[\ts](s)\rbkt{\nodeset}}{\mInf[\ts](s) \rbkt{\nodeset}} = \frac{\mR[\ts](s) \rbkt{\nodeset}}{\abs{\nodeset}} ,
    \label{eq:R_r_relation_rewrite}
\end{equation}
which shows $\mr[\ts](s) \rbkt{\nodeset}$ is the density of the $\influenced$ nodes.
\lemref{the:exact_r} gives the relationship what $\mr[\ts](s) \rbkt{\nodeset}$ satisfies.
\lemm[Equation of $\mr[\ts](s) \rbkt{\nodeset}$]{
    \label{the:exact_r}
    Given the adjacency matrix $\adj(ij)$, conditional activation probability $\infprob{ij}$, the basic reproduction kernel $\repknl : \nodeset \times \sigfd \rightarrow \bbkt{0, \infty}$, and $\repknl[\top]$ is the dual of $\repknl$ in L-2-space of $\nodeset$ in relation to $\mInf(s)$ ($\Ltwo \rbkt{\mInf(s)}$), let $\mi[0] \rbkt{\nodeset}$ be the Radon--Nikodym derivative of $\mI[0](s)\rbkt{\nodeset}$ at $\ts = 0$, $\mr[\ts](s)\rbkt{\nodeset}$ is the solution of
    \begin{equation}
        \rbkt{1 - \mi[0] \rbkt{\nodeset}} \exp \rbkt{-\repknl[\top] \rbkt{s, \nodeset} \cdot \mr[\ts](s) \rbkt{\nodeset}} = 1 - \mr[\ts](s) \rbkt{\nodeset} ,
        \label{eq:r_inequality}
    \end{equation}
    and for $\forall \source \in \nodeset$ and $\forall \randsubset \in \sigfd$, the equation $\repknl \rbkt{s, \randsubset} = \recovery(s)  \sum_{i \in \randsubset} \adj(si) \cdot \infprob{si} \cdot \mInf(s) \rbkt{i}$ holds.
}
\pf{
    SIR model can be represented by the Radon--Nikodym derivatives $\ms[\ts](s)\rbkt{\cdot}$, $\mi[\ts](s) \rbkt{\cdot}$ and $\mr[\ts](s) \rbkt{\cdot}$ as
    \begin{equation}
        \eqlcbkt{
            \begin{aligned}
                \displaystyle \frac{\dif}{\dif \ts} \ms[\ts](s) \rbkt{j} = & - \sum_{i \in \nodeset} \adj(ij) \cdot \infprob{ij} \cdot \ms[\ts](s) \rbkt{j} \mi[\ts](s) \rbkt{i} \mInf(s) \rbkt{i} \\
                \displaystyle \frac{\dif}{\dif \ts} \mi[\ts](s) \rbkt{j} = & \sum_{i \in \nodeset} \adj(ij) \cdot \infprob{ij} \cdot \ms[\ts](s) \rbkt{j} \mi[\ts](s) \rbkt{i} \mInf(s) \rbkt{i}   \\
                \displaystyle                                              & - \frac{1}{\recovery(j) } \mi[\ts](s) \rbkt{j}                                                                        \\
                \displaystyle \frac{\dif}{\dif \ts} \mr[\ts](s) \rbkt{j} = & \frac{1}{\recovery(j) } \mi[\ts](s) \rbkt{j} \period
            \end{aligned}
        }
        \label{eq:density_sir}
    \end{equation}
    By inserting the third equation into the first one, the latter is transformed into
    \begin{equation}
        \displaystyle \frac{\dif}{\dif \ts} \ms[\ts](s) \rbkt{j} = - \sum_{i \in \nodeset} \adj(ij) \cdot \infprob{ij} \cdot \ms[\ts](s) \rbkt{j} \frac{\dif}{\dif \ts} \mr[\ts](s) \rbkt{i} \recovery(i) \mInf(s) \rbkt{i} \period
    \end{equation}
    Suppose $\ms[\ts](s) \rbkt{j}>0$,
    \begin{equation}
        \displaystyle \frac{d}{\ms[\ts](s) \rbkt{j}} \ms[\ts](s) \rbkt{j} = - dt \sum_{i \in \nodeset} \adj(ij) \cdot \infprob{ij} \cdot \frac{\dif}{\dif \ts} \mr[\ts](s) \rbkt{i} \recovery(i) \mInf(s) \rbkt{i} \period
    \end{equation}
    Integrating both sides with respect to $t$, we can obtain
    \begin{equation}
        \displaystyle \ln \ms[u](s) \rbkt{j} - \ln \ms[0](s) \rbkt{j} = - \sum_{i \in \nodeset} \adj(ij) \cdot \infprob{ij} \cdot \frac{\dif}{\dif \ts} \mr[\ts](s)\rbkt{i} \recovery(i) \mInf(s) \rbkt{i} \period
    \end{equation}
    According to Fubini theory~\cite{stein2009real} to change the order of integration, and note that $\mr[0]\rbkt{\cdot} \equiv 0$, we get
    \begin{equation}
        \displaystyle \ln \ms[u](s) \rbkt{j} - \ln \ms[0](s) \rbkt{j} = - \sum_{i \in \nodeset} \adj(ij) \cdot \infprob{ij} \cdot \mr[u](s) \rbkt{i} \recovery(i) \mInf(s) \rbkt{i} \period
        \label{eq:app_change_order}
    \end{equation}
    By \eqaref{eq:density_sir}, $\ms[u](s) \rbkt{j}$ and $\mr[u](s)\rbkt{i}$ are the bounded and convergent function of $u$, so
    \begin{equation}
        \displaystyle \ms[\infty] \rbkt{j} = \lim_{u \rightarrow \infty} \ms[u](s) \rbkt{j}, \quad \mr[\infty](s)\rbkt{i} = \lim_{u \rightarrow \infty} \mr[u](s)\rbkt{i} \period
    \end{equation}
    Let $\Lone(\mInf(s))$ and $\Ltwo(\mInf(s))$ be the L-1 and L-2 space of $\nodeset$ in relation to $\mInf(s)$.
    Then, $0 \leq \mr[u](s) \rbkt{i} \leq 1$, and $\recovery(i) \in \Ltwo(\mInf(s)) \subseteq \Lone(\mInf(s))$.
    According to the dominated convergence theorem, we let $u \rightarrow \infty$ and get
    \begin{equation}
        \displaystyle \ln \ms[\infty](s) \rbkt{j} - \ln \ms[0] \rbkt{j} = - \sum_{i \in \nodeset} \adj(ij) \cdot \infprob{ij} \cdot \mr[\infty](s) \rbkt{i} \recovery(i) \mInf(s) \rbkt{i} \period
    \end{equation}
    We can easily find $\mr[\infty](s) \rbkt{i} \in \Ltwo(\mInf(s))$.
    Note that $\repknl$ can be a linear operator on $\Ltwo(\mInf(s))$ and
    \begin{equation}
        \displaystyle \repknl \mf \rbkt{i} = \recovery(i) \sum_{j \in \nodeset} \adj(ij) \cdot \infprob{ij} \cdot \mf \rbkt{j} \mInf(s) \rbkt{j},
    \end{equation}
    where $\mf$ is any measurable function on $\sigfd$.
    The dual of $\repknl$ on $\Ltwo(\mInf(s))$ is $\repknl[\top]$
    \begin{equation}
        \displaystyle \repknl[\top] \mf \rbkt{j} = \sum_{i \in \nodeset} \recovery(i) \cdot \adj(ij) \cdot \infprob{ij} \cdot \mf \rbkt{i} \mInf(s) \rbkt{i} ,
    \end{equation}
    and then we have
    \begin{equation}
        \displaystyle \ln \ms[\infty](s) \rbkt{j} - \ln \ms[0](s) \rbkt{j} = - \repknl[\top] \mr[\infty](s) \rbkt{\cdot} \period
        \label{eq:app_in}
    \end{equation}
    On the other hand, since $0 \leq \ms[u](s)\rbkt{i} \leq 1$, $\forall \randsubset \in \sigfd$, we have
    \begin{equation}
        \displaystyle \mS[u](s) \rbkt{\randsubset} = \sum_{i \in \randsubset} \ms[u](s) \rbkt{i} \mInf(s) \rbkt{i} .
    \end{equation}
    Let $u \rightarrow \infty$ and using the dominated convergence theorem, we get
    \begin{equation}
        \displaystyle \mS[\infty](s) \rbkt{\randsubset} = \sum_{i \in \randsubset} \ms[\infty](s)\rbkt{i} \mInf(s) \rbkt{i}, \forall \randsubset \in \sigfd .
    \end{equation}
    Thus, $\ms[\infty](s)\rbkt{\cdot}$ and $\mr[\infty](s)\rbkt{\cdot}$ are the Radon--Nikodym derivative of $\mS[\infty](s)\rbkt{\cdot}$ and $\mR[\infty](s)\rbkt{\cdot}$ respectively.
    Given $\mI[0](s)\rbkt{\nodeset} = 0$, $\mS[0](s)\rbkt{\nodeset} + \mI[0](s)\rbkt{\nodeset} = \mInf(s)\rbkt{\nodeset}$, and $\mS[\infty](s)\rbkt{\nodeset} + \mR[\infty](s)\rbkt{\nodeset} = \mInf(s)\rbkt{\nodeset}$, thus $\ms[0](s)\rbkt{\nodeset} = 1 - \mi[0](s)\rbkt{\nodeset}$, $\ms[\infty](s)\rbkt{\nodeset} = 1 - \mr[\ts](s)\rbkt{\nodeset}$, and $\mr[\infty](s) = \mr[\ts](s)\rbkt{\nodeset}$.
    Substitute it into \eqaref{eq:app_in}, we get
    \begin{equation}
        \displaystyle \ln \rbkt{1 - \mr[\ts](s)\rbkt{\nodeset}} - \ln \rbkt{1 - i_0 \rbkt{\nodeset}} = - \repknl[\top] \mr[\ts](s)\rbkt{\nodeset} \period
    \end{equation}
    Taking the exponential on both sides, we can obtain \eqaref{eq:r_inequality}.
    Similar proofs can be found~\cite{guo2011stationary}.
}
For the single source $\source$, $\mI[0](s)\rbkt{\nodeset} = 1$ which leads to its density $\mi[0](s)\rbkt{\nodeset} = \frac{1}{\left| \nodeset \right|} \approx 0$.
Thus, let $\randsubset \rightarrow \nodeset$, $\mr[\ts](s)\rbkt{\nodeset}$ will be the solution of
\begin{equation}
    \exp \rbkt{- \repknl[\top] \rbkt{s, \nodeset} \cdot \mr[\ts](s)\rbkt{\nodeset}} = 1 - \mr[\ts](s)\rbkt{\nodeset} \period
    \label{eq:r_inequality_simplified}
\end{equation}
Next, based on \lemrefs{the:radon_nikodym_theorem}{the:exact_r}, we give the proof of \cororef{the:H_out_approx}.
\pf{
    Given uniformly distributed $\infprob{s \ast} = \uniprob$, then $\repknl[\top] \rbkt{s, \nodeset} = \hat{\recovery} \cdot \uniprob \sum_{i \in \nodeset} \adj(si) \mInf(s) \rbkt{i}$, thus \eqaref{eq:r_inequality_simplified} can be rewritten as
    \begin{equation}
        \exp (- \hat{\recovery} \cdot \uniprob \cdot \eqabs{ \neighbor[1](s) } \cdot \mr[\tsp](s)\rbkt{\nodeset}) = 1 -\mr[\tsp](s)\rbkt{\nodeset} ,
    \end{equation}
    where there is not closed form solution for $\mr[\tsp](s)\rbkt{\nodeset}$, but it is determined by the constant value $\hat{\recovery} \cdot \uniprob \cdot \abs{  \neighbor[1](s) }$.
    Once $\infprob{s \ast} = p$ is constant, we could use a proper $\nborder$ to approximate $\mr[\tsp](s)\rbkt{\nodeset}$ by $\frac{1}{\abs{ \nodeset }}\sum_{k=0}^n \abs{  \neighbor[k](s) }$, since the influence will only propagate on the edges to the neighbors.
    That means $\mR[\tsp](s)\rbkt{\nodeset} = \sum_{k=0}^{\nborder} \abs{ \neighbor[k](s) }$ and $\hout(s)$ will be nodes in $\cup_{k=0}^{\nborder} \neighbor[k](s)$.
    This gives us the solution for \cororef{the:H_out_approx}.
}

\section{Proof of \texorpdfstring{\theref{the:H_in}}{Theorem 2}}\label{sec:proof-of-theorem-Hin}
\pf{
    Given the number of \noticed nodes $\mU[\tsp](s)\rbkt{\nodeset} = \mR[\tsp](s)\rbkt{\nodeset} \cdot \Qt$.
    Based on the proof of \theref{the:H_out}, the node set that includes \noticed nodes will be $\eqcbkt{ \nodei: \sum_{\singlepath \in \paths(si)} \prod_{\edge(ab) \in \singlepath} \infprob{ab} \ge \frac{1}{\mR[\tsp](s)\rbkt{\nodeset} \rbkt{\Qt + 1}} }$.
    In addition, the \noticed nodes also satisfy $\eqcbkt{ \nodei: \sum_{\singlepath \in \paths(si)} \prod_{\edge(ab) \in \singlepath} \infprob{ab} < \frac{1}{\mR[\tsp](s)\rbkt{\nodeset}} }$, since they are not the nodes of outward influence propagation.
    Thus, the expected \noticed node set $\widehat{\noticedset(s)}$ will be $\eqcbkt{ \nodei: \frac{1}{\mR[\tsp](s)\rbkt{\nodeset} \rbkt{\Qt + 1}} \le \sum_{\singlepath \in \paths(si)} \prod_{\edge(ab) \in \singlepath} \infprob{ab} < \frac{1}{\mR[\tsp](s)\rbkt{\nodeset}} }$.
    Again, $\lim_{k \rightarrow \infty} \E \eqbbkt{ \eqabs{\noticedset(s) \setminus \widehat{\noticedset(s)}} } = 0$.
    Hence $\noticedset(s)$ will be converged in probability to $\widehat{\noticedset(s)}$.
    Once we have the $\noticedset(s)$, the $\hin(s)$ will be the nodes from the paths of response in $\noticedset(s)$, which will be $\bigcup_{j \in \border(s), i \in \noticedset(s)} \eqcbkt{\node(k): \node(k) \in \paths(ji) ; \node(k) \ne \nodej ; \adj(is) \ne 0}$.
}

\section{Proof of \texorpdfstring{\lemref{the:complexity_of_GNNs}}{Lemma 1}}\label{sec:proof-of-lemma-gnn-comlexity}
\pf{
For the time complexity of training, a GNN needs to compute the embeddings of all the nodes in a mini-batch for $\recurrence$ times due to the $\recurrence$ layers of the GNN.
Since each node embedding update involves a matrix multiplication $\nodefeat[(\layers)]\netweight[(\layers)]$ with $\netweight[(\layers)]\in\mathbb{R}^{\realdim\times\realdim}$, the cost per node per layer is $\bigO\rbkt{\realdim[2]}$.
With graph sparsity $\gsparsity$, the $k$th hop contributes on average $\gsparsity[k]$ nodes, so the total node count in an $\nborder$-order sub-graph is $\sum_{i=0}^{\nborder}\gsparsity[i]$.
Therefore, we obtain the training time complexity of GNNs as $\bigO \rbkt{ \ngbat \cdot \sum_{i=0}^{\nborder} \gsparsity[i] \cdot \realdim[2] \cdot \recurrence}$.
}

\section{Proof of \texorpdfstring{\theref{the:ve_speedup}}{Theorem 3}}\label{sec:proof-of-speedup}
\pf{
The time complexity of GNN training with and without virtual edges is $\bigO \rbkt{\ngbat \cdot \sum_{i=0}^{\nborder} \gsparsity[i] \cdot \realdim[2] \cdot \recurrence}$ and $\bigO \rbkt{\ngbat \cdot \sum_{i=0}^{\nborder \cdot \rbkt{\Q + 1}} \gsparsity[i] \cdot \realdim[2] \cdot \recurrence}$ respectively, so using virtual edges will be $\frac{\ngbat \cdot \sum_{i=0}^{\nborder \cdot \rbkt{\Q + 1}} \gsparsity[i] \cdot \realdim[2] \cdot \recurrence}{\ngbat \cdot \sum_{i=0}^{\nborder} \gsparsity[i] \cdot \realdim[2] \cdot \recurrence} = \frac{\sum_{i=0}^{\nborder \cdot \rbkt{\Q + 1}} \gsparsity[i]}{\sum_{i=0}^{\nborder} \gsparsity[i]}$ times faster.
}

\end{document}